\documentclass[12pt]{article}
\pdfoutput=1
\usepackage{putex}
\usepackage{amsmath}
\usepackage{amssymb}
\usepackage{epsf}
\usepackage{tikz}
\usepackage{graphicx}
\usepackage{caption}
\usepackage{subcaption}
\usepackage{epstopdf}
\usepackage{enumerate}
\usepackage{cite}
\usepackage{tensor}
\usepackage{slashed}
\usepackage{feynmf}
\usepackage{hyperref}
\usetikzlibrary{decorations.markings}
\usetikzlibrary{decorations.pathmorphing}
\newlength{\lx}
\newlength{\ly}
\newlength{\lxx}
\newlength{\radius}
\newlength{\lyy}
\numberwithin{equation}{section}

\renewcommand{\baselinestretch}{1.2}

\newcommand{\be}{\begin{equation}}
\newcommand{\ee}{\end{equation}}
\newcommand{\bes}{\begin{equation*}}
\newcommand{\ees}{\end{equation*}}
\newcommand{\bea}{\begin{eqnarray}}
\newcommand{\eea}{\end{eqnarray}}
\newcommand{\beas}{\begin{eqnarray*}}
\newcommand{\eeas}{\end{eqnarray*}}

\newcommand{\p}{\partial}
\newcommand{\bmat}{\begin{bmatrix}}
\newcommand{\emat}{\end{bmatrix}}

\def\wb{\overline{w}}

\def\zb{\overline{z}}

\def\hb{\overline{h}}

\def\rt{\rightarrow}

\newcommand{\lr}[1]{\left( #1 \right)}
\newcommand{\Farg}[3]{\left( \begin{array}{c} #1 \\ #2 \end{array} \Big| #3 \right)}
\DeclareMathOperator{\free}{free}
\DeclareMathOperator{\geo}{geo}
\DeclareMathOperator{\arctanh}{arctanh}

\def\cW{{\cal W}}
\def\vphi{\varphi}
\def\cG{{\cal G}}
\def\F{{\cal F}}

\def\Oc{{\cal O}}
\def\vs{\vskip .1 in}
\def\D{\Delta}

\def\g{\gamma}
\def\a{\alpha}

\newcommand{\e}[2] {\begin{equation} \label{#1} #2 \end{equation}}
\newcommand{\es}[2] {\begin{equation} \label{#1} \begin{split} #2 \end{split} \end{equation}}

\def\wb{\overline{w}}

\def\sec{\section}
\def\subsec{\subsection}

\def\eqr{\eqref}

\def\l{\lambda}

\def\rar{\rightarrow}

 \newmuskip\pFqmuskip

\newcommand*\pFq[6][8]{%
  \begingroup % only local assignments
  \pFqmuskip=#1mu\relax
  \mathcode`\,=\string"8000
  \begingroup\lccode`\~=`\,
  \lowercase{\endgroup\let~}\pFqcomma
  {}_{#2}F_{#3}{\left[\genfrac..{0pt}{}{#4}{#5};#6\right]}%
  \endgroup
}
\newcommand{\pFqcomma}{\mskip\pFqmuskip}
\definecolor{propagatorblue}{RGB}{0,112,192}

\begin{document}

\tikzset{->-/.style={decoration={
markings,
mark=at position #1 with {\arrow{>}}},postaction={decorate}}}
\tikzset{point/.style={insert path={ node[scale=2.5*sqrt(\pgflinewidth)]{.} }}}
\tikzset{->-/.style={decoration={
markings,
mark=at position #1 with {\arrow{>}}},postaction={decorate}}}
\tikzset{-dot-/.style={decoration={
markings,
mark=at position #1 with {\draw[fill=black,color=black] circle [radius=2pt,fill=blue];}},postaction={decorate}}}

\tikzset{-dot2-/.style={decoration={
markings,
mark=at position #1 with {\draw[fill=black,color=black] circle [radius=1.5pt,fill=blue];}},postaction={decorate}}}

 %%% in this line added a ;
\tikzset{snake it/.style={decorate, decoration=snake}}

\institution{UCLA}{Department of Physics and Astronomy, University of California, Los Angeles, CA 90095, USA}

\institution{PU}{Department of Physics, Princeton University, Princeton, NJ 08544, USA}

\title{Semiclassical Virasoro Blocks from\\AdS$_3$ Gravity}

\authors{Eliot Hijano\worksat{\UCLA}, Per Kraus\worksat{\UCLA}, Eric Perlmutter\worksat{\PU}, River Snively\worksat{\UCLA}}

\abstract{We present a unified framework for the holographic computation of Virasoro conformal blocks at large central charge. In particular, we provide bulk constructions that correctly reproduce all semiclassical Virasoro blocks that are known explicitly from conformal field theory computations. The results revolve around the use of geodesic Witten diagrams, recently introduced in \cite{Hijano:2015zsa}, evaluated in locally AdS$_3$ geometries generated by backreaction of heavy operators. We also provide an alternative computation of the heavy-light semiclassical block -- in which two external operators become parametrically heavy -- as a certain scattering process involving higher spin gauge fields in AdS$_3$; this approach highlights the chiral nature of Virasoro blocks. These techniques may be systematically extended to compute corrections to these blocks and to interpolate amongst the different semiclassical regimes.}

\date{}

\maketitle
\setcounter{tocdepth}{2}
\tableofcontents

\renewcommand{\baselinestretch}{1.2}

\sec{Introduction}

Correlation functions in conformal field theories admit a decomposition in terms of conformal blocks, obtained by using the OPE to reduce products of local operators at distinct points to a sum of local operators at a single point, and collecting the contribution of operators lying in a single representation of the conformal algebra; see e.g. \cite{Ferrara:1971vh,Ferrara:1973vz,Ferrara:1974ny,Belavin:1984vu,DiFrancesco:1997nk,Dolan:2000ut,Dolan:2003hv,Costa:2011dw}.
This yields a concrete algorithm to go from the basic CFT data --- a list of primary operators and their OPE coefficients --- to correlation functions. The conformal blocks are fully determined by conformal symmetry, and so are universal to all CFTs. They feature prominently in many applications of CFT, including in the conformal bootstrap program \cite{Rattazzi:2008pe,ElShowk:2012ht} and in the study of the emergence of bulk locality from CFT \cite{Heemskerk:2009pn,Heemskerk:2010ty,Penedones:2010ue}.

Given a consistent theory of gravity in AdS, one can compute correlation functions that obey CFT axioms, and hence admit a decomposition into conformal blocks. A natural question is: what object in AdS gravity computes a CFT conformal block? In previous work \cite{Hijano:2015zsa} we answered this question for AdS$_{d+1}$/CFT$_d$ for any $d$. There, an elegant prescription was found in the case of four-point conformal blocks with external scalar operators. The main result is that a conformal block --- more precisely, a conformal partial wave --- is obtained from a ``geodesic Witten diagram.'' This is essentially an ordinary exchange Witten diagram but with vertices integrated over geodesics connecting the external operators, rather than over all of AdS. See Figure \ref{f1}. The case of $d=2$ is special because the global conformal algebra is enhanced to two copies of the Virasoro algebra. The corresponding Virasoro conformal blocks are much richer objects, containing an infinite number of global conformal blocks. In the present work we address the bulk construction of the Virasoro blocks.

Unlike the case of global blocks, Virasoro blocks depend on the central charge $c$ of the theory. Because we will be working in the context of classical gravity, and $1/c$ plays the role of $\hbar$ in the bulk, we must restrict attention to the regime $c \rt \infty$, corresponding to so-called semiclassical Virasoro blocks. There are various ways to take this limit, corresponding to the manner in which the operator dimensions behave as $c\rt \infty$. Two natural choices bookend the spectrum of possibilities: either keep all operator dimensions fixed, or let all operator dimensions scale linearly with $c$. As we review in Section \ref{review}, analytical expressions for the Virasoro blocks have been derived at various points on this spectrum using CFT techniques.  Indeed, with a few exceptions \cite{Belavin:1984vu,zamo:ashkin, Gamayun:2012ma}, these are some of the only analytical expressions for Virasoro blocks available.

\begin{figure}[t!]
   \begin{center}
 \includegraphics[width = .76\textwidth]{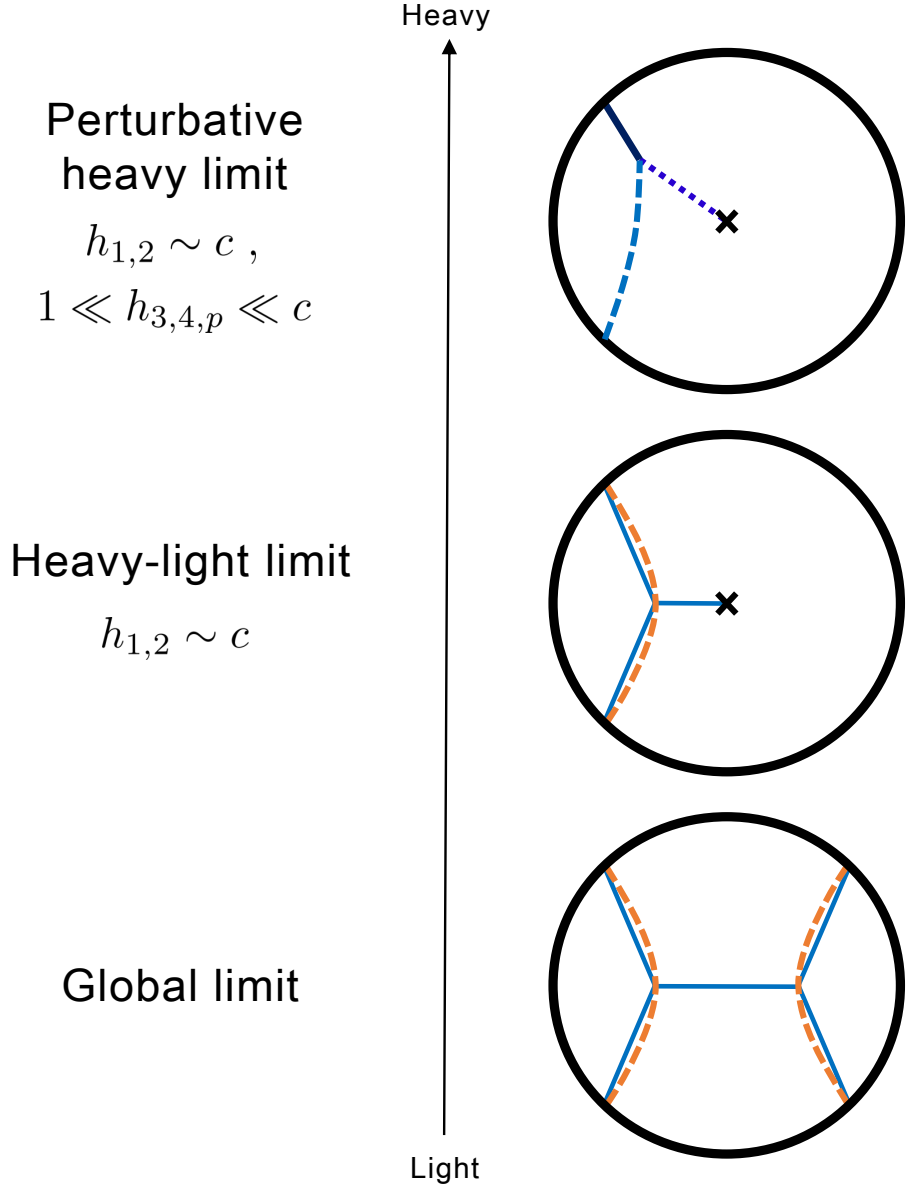}
 \caption{The spectrum of gravity duals of large $c$ Virasoro blocks. Operator dimensions increase from bottom to top; $h_i$ and $h_p$ denote external and internal holomorphic operator dimensions, respectively. In the limit of fixed dimensions, the Virasoro block becomes the global block, represented by a geodesic Witten diagram. Upon ramping up two external dimensions to enter the heavy-light regime, the bulk dual becomes a geodesic Witten diagram evaluated in a conical defect geometry. Further taking the remaining dimensions to scale with $c$, albeit perturbatively, one minimizes the worldline action of a cubic vertex of geodesics in the presence of the defect. This is equivalent to making a saddle-point approximation to the heavy-light geodesic Witten diagram. Not shown is the fully non-perturbative Virasoro block for all heavy operators, whose form is unknown.}
 \label{f1a}
 \end{center}
 \end{figure}

In what follows, we will present a framework that computes all known semiclassical Virasoro blocks using 3D gravity. Partial results on bulk derivations of Virasoro blocks are already known \cite{Fitzpatrick:2014vua,Hijano:2015rla,Alkalaev:2015wia}, and we will incorporate and reproduce them here. One object whose bulk dual has not been constructed as yet is the elegant formula obtained recently by Fitzpatrick, Kaplan and Walters (FKW) \cite{Fitzpatrick:2015zha}, for the four-point conformal block in the case that two external operator dimensions grow linearly in $c$, while the rest remain fixed (see (\ref{FKWresult}) and \eqr{FKWresult2}). This is known as the ``heavy-light limit.'' By combining the ideas of \cite{Fitzpatrick:2014vua,Hijano:2015rla,Alkalaev:2015wia} with our other work on global blocks \cite{Hijano:2015zsa}, we will indeed arrive at a more complete story for the holographic construction of semiclassical Virasoro blocks. We provide a diagrammatic overview in Figure \ref{f1a}.

Our main results can be summarized as follows. First, it is well-known that in the $c\rar\infty$ limit with operator dimensions held fixed, the Virasoro block reduces to the global block \cite{zamo,Zamo:lectures}. Therefore, the geodesic Witten diagram provides the bulk construction of the Virasoro block in this simple limit. More significantly, we will reproduce the FKW result in the heavy-light limit. The main idea is essentially to start with the geodesic Witten diagram for the global block, and allow one of the geodesics to backreact on AdS$_3$.
This sets up a conical defect or BTZ geometry for the remaining part of the geodesic Witten diagram corresponding to the light operators; explicit computation leads quickly to the correct result. (The appearance of a defect or black hole depends on whether the heavy operator dimension is above or below the black hole threshold, $h=c/24$.) Interpolation amongst the various semiclassical limits may be systematically computed, in principle, by treating backreaction effects, providing an intuitive bridge between the different regimes. We will explicitly demonstrate in Appendix \ref{worldapp}, for instance, that the saddle-point approximation to the geodesic Witten diagram for the heavy-light block manifestly reduces to the worldline prescription for the perturbative heavy blocks given in \cite{Hijano:2015rla}. This correctly interpolates between the two regimes. We have therefore provided a bulk construction for all known semiclassical Virasoro blocks.

We will in fact provide two distinct but complementary constructions of the heavy-light blocks. The first, just described, employs scalar fields propagating in the background of a locally AdS$_3$ conical defect geometry. The second takes advantage of the fact that the Virasoro blocks are chiral objects. Accordingly, we can use chiral currents to represent the operators; the analytic continuation from integer operator dimensions (i.e. spins) to arbitrary dimensions is the trivial one, since the blocks' dependence on dimensions is rational. These currents are dual to massless higher spin gauge fields living in AdS$_3$. With efficient use of higher spin gauge transformations, we are able to reproduce the semiclassical blocks in a theory of higher spin gravity in AdS$_3$.

The remainder of this note is organized as follows. In section \ref{review} we review some basic facts and results about semiclassical Virasoro blocks. In section \ref{scalar} we show how a scalar field computation using geodesic Witten diagrams reproduces the result of FKW. The alternative higher spin approach is presented in section \ref{higherspin}. Appendices \ref{integral} and \ref{ap:Hsap} contain some technical results needed in the main body of the paper, and Appendix \ref{worldapp} explains the relation of this work to the approach in \cite{Hijano:2015rla}.

%%%%%%%%%%%%%%%%%%%%%%%%%%%%%%%%%%%%%%%%%%%%%%%%%%%%%%%%%%%%%%%%%%%%%%%%%%%%%%%%
\section{Review of semiclassical Virasoro blocks}
\label{review}
%%%%%%%%%%%%%%%%%%%%%%%%%%%%%%%%%%%%%%%%%%%%%%%%%%%%%%%%%%%%%%%%%%%%%%%%%%%%%%%%

We consider a four-point function of Virasoro primary operators $\Oc_i(z_i,\zb_i)$ on the plane,
\be
\langle \Oc_1(z_1,\zb_1)\Oc_2(z_2,\zb_2)\Oc_3(z_3,\zb_3)\Oc_4(z_4,\zb_4)\rangle \text{ .}
\ee
$\Oc_i$ has holomorphic and anti-holomorphic conformal weights $(h_i,\hb_i)$, respectively. Using SL(2,$\mathbb{C}$) invariance, three of the operators can be taken to specified locations. It will be convenient to thereby consider
\be
\langle \Oc_1(\infty,\infty)\Oc_2(0,0)\Oc_3(z,\zb)\Oc_4(1,1)\rangle  \text{ ,}
\ee
where $\Oc_1(\infty,\infty) = \lim_{z_1, \zb_1 \rt \infty} z_1^{2h_1} \zb_1^{2\hb_1}\Oc_1(z_1,\zb_1)$ inside the correlator. A basis for the Hilbert space of the CFT consists of the set of primary states $|\Oc_p\rangle$ (equivalently, local primary operators $\Oc_p$) and their Virasoro descendants, i.e. the set of irreducible highest weight representations of the Virasoro algebra. This implies the existence of a Virasoro conformal block decomposition of the four-point function,
\e{}{\langle \Oc_1(\infty,\infty)\Oc_2(0,0)\Oc_3(z,\zb)\Oc_4(1,1)\rangle = \sum_pC_{12p}C^p_{~34} \F(h_i,h_p,c;z-1)\overline{\F}(\hb_i,\hb_p,c;\zb-1)  \text{ ,}}
where the sum runs over all irreducible representations of the Hilbert space. We use $h_i$ to stand for $h_{1,2,3,4}$. For simplicity, we have assumed equal left- and right-moving central charges. The fact that the holomorphic and anti-holomorphic Virasoro algebras commute with each other leads to holomorphic factorization for given $p$.

The Virasoro blocks can be conveniently defined using a projector, which we denote $P_p$, acting within the Hilbert space. The $s$-channel Virasoro block is obtained by inserting this projector between the operators $\Oc_2$ and $\Oc_3$:\footnote{We won't concern ourselves with the normalization of this function, which is fixed by matching its small $z$ behavior to the $\Oc_1\Oc_2$ and $\Oc_3\Oc_4$ OPEs, and throughout will freely discard any $z$-independent prefactors.}
\be
\langle \Oc_1(\infty,\infty)\Oc_2(0,0)\,P_p\, \Oc_3(z,\zb)\Oc_4(1,1)\rangle = \F(h_i,h_p,c;z-1)\overline{\F}(\hb_i,\hb_p,c;\zb-1) ~.
\label{VVbar}
\ee
We refer to $\F(h_i,h_p,c;z-1)$ alone as the Virasoro block.\footnote{In $d$-dimensional conventions, as in \cite{Hijano:2015zsa}, this projection is better known as a conformal partial wave. However, in 2d CFT literature, one often finds the convention used here, in which the $z\rar 1$ expansion of the block itself starts at $(z-1)^{h_p-h_3-h_4}$, as opposed to $(z-1)^{h_p}$.}

Unlike for global conformal blocks, no closed-form expressions for Virasoro blocks are known, except in some very special cases \cite{Belavin:1984vu,zamo:ashkin, Gamayun:2012ma}. We briefly mention what is known in general. OPE considerations reveal that $\F(h_i,h_p,c;z)$ has the structure $z^{h_p-h_3-h_4}f(z)$, where $f(z)$ is analytic in the unit disk. Zamolodchikov \cite{zamo, Zamo:lectures} has provided recursion relations allowing one to efficiently compute terms in the power series expansion of $f(z)$ around the origin. These recursion relations can be solved \cite{Perlmutter:2015iya, Ribault:2014hia}. The expansion coefficients are rational functions of the conformal weights, which rapidly become extremely complicated. The coefficients have also been computed using combinatorial methods inspired by the AGT correspondence \cite{Alday:2009aq,Alba:2010qc}.

Of greater relevance here is the semiclassical limit corresponding to taking $c\rt \infty$. If $h_i$ are all held fixed in the limit, the Virasoro block simply reduces to the global block, which is a hypergeometric function \cite{Ferrara:1974ny}:
\e{}{\lim_{c\rar\infty} {\cal F}(h_i,h_p,c;z-1) =
(z-1)^{h_p-h_3-h_4}{}_2F_1(h_p-h_{12},h_p+h_{34}; 2h_p; z-1)~.}
where $h_{ij}\equiv h_i-h_j$. Instead, we are interested in the case in which we hold fixed some ratios $h_i/c$. If all ratios $h_i/c$ are held fixed in the limit, then one can apply Zamolodchikov's monodromy method (well reviewed in \cite{Harlow:2011ny,Fitzpatrick:2014vua}) to determine the semiclassical Virasoro block. The equations resulting from this approach turn out to be equivalent to those of 3D gravity with negative cosmological constant; this becomes especially transparent in the Chern-Simons formulation (see e.g. \cite{Hijano:2015rla}). However, this is still too complicated to admit an exact solution. Progress can be made in perturbation theory by taking $h_3/c, h_4/c, h_p/c \ll 1$, keeping $h_1/c$ and $h_2/c$ finite. Results obtained in this approach can be found in \cite{Fitzpatrick:2014vua,Hijano:2015rla,Alkalaev:2015wia}. The 3D gravity picture in this case consists, at lowest order in the above small parameters, of particle worldlines moving in a background geometry of the ``heavy" operators $h_{1,2}$. Higher orders in perturbation theory account for the backreaction of the particles on the geometry. We called this the perturbative heavy limit in Figure \ref{f1a}.

Let us give slightly more detail. To distinguish heavy and light operators we now write
\be
h_1= h_{H_1}~,~~h_2= h_{H_2}~,~~h_3= h_{L_1}~,~~h_4= h_{L_2}  \text{ .}
\ee
The bulk prescription for computing the semiclassical Virasoro block to first order\footnote{This regime can also be described as holding fixed $h_{L_1,L_2,p}$, and then working to first order in $1/h_{L_1,L_2,p}$. These two procedures turn out to agree, as discussed in \cite{Fitzpatrick:2015zha}.} in $h_{L_1}/c, h_{L_2}/c, h_p/c \ll 1$ was first explained in \cite{Fitzpatrick:2014vua} in the simplified case of $h_{L_1}=h_{L_2}, h_{H_1}=h_{H_2}, h_p=0$, which corresponds to the vacuum Virasoro block. The heavy operators backreact to generate the  metric
\be
ds^2= {\alpha^2 \over \cos^2\rho}\left( {d\rho^2\over \alpha^2}+d\tau^2 +\sin^2\rho \,d\phi^2\right) \text{ ,}
\label{condef}
\ee
with $\phi\cong \phi+2\pi$. For real $\alpha<1$, this is a conical defect solution with a singularity at $\rho=0$; for $\alpha^2<0$ it becomes a BTZ black hole after Wick rotation.  This can be thought of as representing the geometry sourced by a particle of mass $m^2=4h_{H_1}(h_{H_1}-1)$ sitting at the origin of global AdS$_3$, where
\e{alpha}{\alpha=\sqrt{1- {24 h_{H_{1}}\over c}}~.}
The ``light'' operators are incorporated by a geodesic in the background \eqr{condef} connecting their locations on the boundary. The appearance of geodesics makes sense because these operators, while parametrically lighter than the heavy operators, still have $h_{L_1}/c, h_{L_2}/c$ fixed in the large $c$ limit. The Virasoro vacuum block is then simply given by $e^{-mL}$, where $m^2=4h_{L_1}(h_{L_1}-1)$, and $L$ is the geodesic length, regulated with a near boundary cutoff. An elementary computation yields $e^{-mL} \propto \left|\sin{\alpha w\over 2}\right|^{-4h_{L_1}}$, which is the correct result derived from CFT \cite{Fitzpatrick:2014vua}.

In \cite{Hijano:2015rla} this was further generalized to allow for $h_{L_1}\neq h_{L_2}$ and $h_p \neq 0$. The picture is now of three geodesic segments, living in the geometry (\ref{condef}), and joined at a cubic vertex. Two of the geodesics are anchored at the locations of $\Oc_{L_1}$ and $\Oc_{L_2}$, while the geodesic corresponding to $\Oc_p$ stretches between the cubic vertex and the singularity at $\rho=0$. The location of the cubic vertex is found by extremizing the total geodesic action $S=m_pL_p + m_1 L_1 + m_2 L_2$, and then the Virasoro block in this regime is obtained from $e^{-S}$. In \cite{Hijano:2015rla} it was explained why this prescription works, by thinking about the relationship between Zamolodchikov's monodromy method and the linearized backreaction produced by these worldlines.

\subsec{The heavy-light semiclassical limit}
The case considered in the present work corresponds to
\be\label{hllim}
c\rt \infty\quad {\rm with}\quad {h_{H_1}\over c},~~{h_{H_2}\over c},~~ h_{H_1}-h_{H_2},~~ h_{L_1},~~h_{L_2},~~h_p\quad {\rm fixed}~.
\ee
This so-called ``heavy-light limit'' was considered recently in \cite{Fitzpatrick:2015zha}. By a clever use of conformal mappings, they were able to relate the Virasoro block in this limit to a global block, with a result
\e{FKWresult}{\langle \Oc_{H_1}(\infty,\infty)\Oc_{H_2}(0,0)\,P_p \,\Oc_{L_1}(z,\zb)\Oc_{L_2}(1,1)\rangle \rar\F\left(h_i,h_p,c;z-1\right)\overline{\F}(\hb_i,\hb_p,c;\zb-1) \text{ ,}}
with
\e{FKWresult2}{\F\left(h_i,h_p,c;z-1\right)= z^{(\alpha-1)h_{L_1}} (1-z^\alpha)^{h_p-h_{L_1}-h_{L_2}} {_2F_1}\Big(h_p+h_{12},h_p-{H_{12}\over \alpha},2h_p;1-z^\alpha\Big) \text{ ,}}
where $\a$ was defined in \eqr{alpha},
and
\be
h_{12}\equiv h_{L_1}-h_{L_2}~,\quad H_{12}\equiv h_{H_1}-h_{H_2}~.
\ee
Note that in the definition of $\alpha$ it doesn't matter whether $h_{H_1}$ or $h_{H_2}$ appears, since we are taking $(h_{H_1}-h_{H_2})/c \rt 0$ in the limit. Setting $\alpha=1$ yields the global conformal block. The result (\ref{FKWresult}) can be checked by expanding in $z-1$ and matching to the series expansion (up to some finite order).

Our goal in the remainder of this paper is to show how to reproduce this result from AdS$_3$ gravity. To this end, it will also useful to rewrite the result on the cylinder, $z=e^{iw}$, with $w=\phi+i\tau$. Taking into account the usual transformation rule for primary operators, and dropping a constant multiplicative prefactor, we have, in the heavy-light limit,
\es{FKWw}{
&\langle \Oc_{H_1}(\tau=-\infty)\Oc_{H_2}(\tau=\infty)\, P_{p}\, \Oc_{L_1}(w,\wb) \Oc_{L_2}(0,0)\rangle \rar\F(h_i,h_p,c;w)\overline{\F}(\hb_i,\hb_p,c;\wb)~,\\&\\ &\F(h_i,h_p,c;w)=\left(\sin{\alpha w \over 2}\right)^{-2h_{L_1}} \left(1-e^{i\alpha w} \right)^{h_p+h_{12}} {_2{F_1}}\Big(h_p+h_{12},h_p-{H_{12}\over \alpha} ,2h_p;1-e^{i\alpha w} \Big)~.}

%%%%%%%%%%%%%%%%%%%%%%%%%%%%%%%%%%%%%%%%%%%%%%%%%%%%%%%%%%%%%%%%%%%%%%%%%%%%%%%%
\section{Semiclassical Virasoro blocks from AdS$_3$ gravity}
\label{scalar}
%%%%%%%%%%%%%%%%%%%%%%%%%%%%%%%%%%%%%%%%%%%%%%%%%%%%%%%%%%%%%%%%%%%%%%%%%%%%%%%%

\subsection{Bulk prescription}

The bulk recipe for reproducing (\ref{FKWw}) is easy to motivate once we recall some previous results. As reviewed above and made clear in Figure \ref{f1a}, the heavy-light limit \eqr{hllim} sits halfway between two other large $c$ limits: holding all $h_i$ and $h_p$ fixed, or holding ratios $h_i/c$ and $h_p/c$ fixed. While the bulk prescription for computing the Virasoro block in the latter limit was just described in the previous section, the prescription for the former limit may be extracted from more recent work \cite{Hijano:2015zsa}, as we now discuss. We can thus obtain the prescription for computing the heavy-light block as a middle ground between those known results.

Consider setting $\alpha=1$ in \eqr{FKWw}, which as noted above yields the global conformal block. This is equivalent to holding all $h_i$ fixed as $c\rar\infty$. In \cite{Hijano:2015zsa}, a simple bulk setup for computing conformal partial waves for symmetric, traceless spin-$\ell$ exchange was proposed and proven in arbitrary spacetime dimension. The picture is that of a geodesic Witten diagram, as we now explain in the setting of AdS$_3$/CFT$_2$. Consider the global block corresponding to exchange of $\Oc_p$. For simplicity, we take $\Oc_p$ to be spinless, so $h_p=\hb_p\equiv\D/2$. To define the geodesic Witten diagram, we begin with an ordinary exchange Witten diagram in AdS$_3$, where the exchanged field is a scalar of mass $m^2 = \D(\D-2)$.
 \begin{figure}[t!]
   \begin{center}
 \includegraphics[width = 0.6\textwidth]{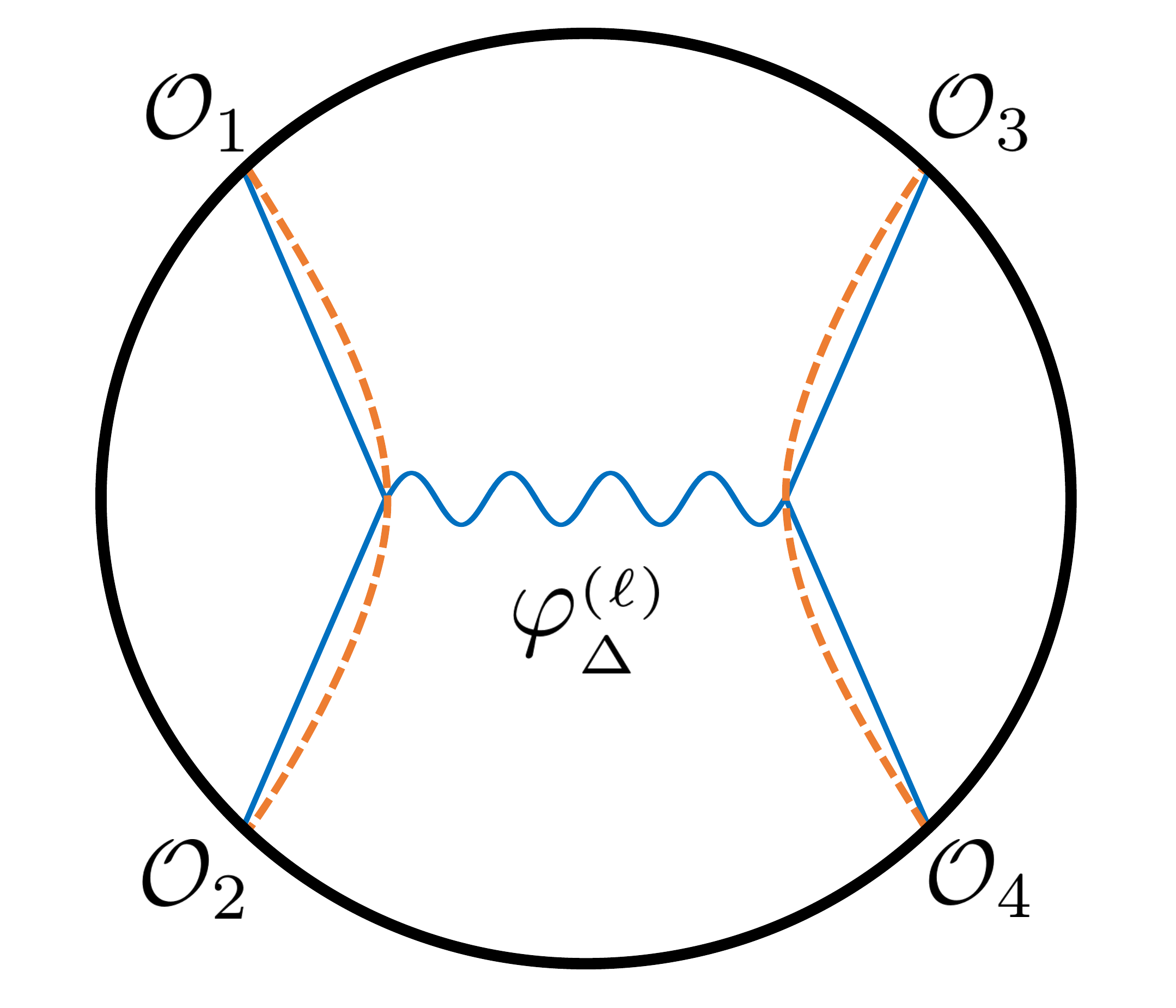}
 \caption{This is a geodesic Witten diagram in AdS$_{d+1}$, for the exchange of a symmetric traceless spin-$\ell$ tensor with $m^2=\D(\D-d)-\ell$ in AdS units,  introduced in \cite{Hijano:2015zsa}. The vertices are integrated over the geodesics connecting the two pairs of boundary points, here drawn as dashed orange lines. This computes the conformal partial wave for the exchange of a CFT$_d$ primary operator of spin $\ell$ and dimension $\D$. When $d=2$, this yields the product of holomorphic and anti-holomorphic global conformal blocks. To compute the heavy-light Virasoro blocks instead, we allow one geodesic to backreact, creating a conical defect.}
 \label{f1}
 \end{center}
 \end{figure}
In the full Witten diagram, the cubic vertices are integrated over all of AdS; to  compute instead the geodesic Witten diagram, and hence the global conformal block, we restrict the integration to the bulk geodesics $\gamma_{12}$ and $\gamma_{34}$ connecting the indicated boundary points. Then the geodesic Witten diagram for scalar exchange, denoted ${\cal W}_{\D,0}$, is
\es{geoblock}{&{\cal W}_{\D,0}(x_i)=\\&
\int_{\gamma_{12}}\!d\lambda \int_{\gamma_{34}}\!d\lambda' G_{b\p}(x_1,y(\lambda))G_{b\p}(x_2,y(\lambda))\times G_{bb}(y(\lambda),y(\lambda');\D)\times
G_{b\p}(x_3,y(\lambda'))G_{b\p}(x_4,y(\lambda')) \text{ ,}}
where $\lambda$ and $\lambda'$ denote proper length. See Figure \ref{f1}. $G_{b\p}$ and $G_{bb}$ are bulk-to-boundary and bulk-to-bulk propagators, respectively. We use the convention that $x$ denotes a point on the boundary, and $y$ a point in the bulk. Up to normalization factors that can be found in \cite{Hijano:2015zsa}, (\ref{geoblock}) is equal to the corresponding product of holomorphic and anti-holomorphic global blocks for $\Oc_p$ exchange. This generalizes nearly verbatim to $d>2$. While it is familiar that geodesics can appear in Witten diagrams as an approximation in the case that the mass of the corresponding field is large, here there is no approximation: (\ref{geoblock}) is an exact expression for fields of any mass, i.e. any operator dimensions.

\begin{figure}
    \centering
    \begin{subfigure}[b]{0.48\linewidth}        %% or \columnwidth
        \centering
       \begin{tikzpicture}[scale=0.55]
\draw[black,very thick] (0,0) -- (\lxx,0);
\draw[black,very thick] (0,0) -- (\lxx/5,\lyy);
\draw[black,very thick] (\lxx,0) -- (\lxx+\lxx/5,\lyy);
\draw[black,very thick] (\lxx/5,\lyy) -- (6\lxx/5,\lyy);
\draw[black,very thick,fill=black](\lxx/4,\lyy/4) circle (\lxx/120);
\draw (\lxx/4,\lyy/4)  node[below=1] { ${\cal O}_{H,1} (0)$};
\draw[black,very thick,fill=black](5\lxx/6,\lyy/4) circle (\lxx/120);
\draw (5\lxx/6,\lyy/4) node[below=1] { ${\cal O}_{H,2} (\infty)$};
\draw[orange,dashed,ultra thick](4\lxx/9,3\lyy/4) arc (180:0:2 and 2) node [pos=0.15,above=8] {\textcolor{black}{}} node [pos=0.85,above=8] {\textcolor{black}{}};
\draw[propagatorblue,very thick] (5.5\lxx/9,23\lyy/20) -- (\lxx/4,7\lyy/5) node [pos=0.4,above=2] {\textcolor{black}{$h_p$}};

\draw[propagatorblue,very thick] (5.5\lxx/9,23\lyy/20) -- (7\lxx/9,3\lyy/4);
\draw[propagatorblue,very thick] (5.5\lxx/9,23\lyy/20) -- (4\lxx/9,3\lyy/4);

\draw[black,very thick,fill=black](4\lxx/9,3\lyy/4) circle (\lxx/120);
\draw (4\lxx/9,3\lyy/4)  node[below=15,left=-10] {${\cal O}_{L,1} (z_1)$};
\draw[black,very thick,fill=black](7\lxx/9,3\lyy/4) circle (\lxx/120);
\draw (7\lxx/9,3\lyy/4)  node[below=15,right=-10] {${\cal O}_{L,2}(z_2)$};
\draw[black,very thick,dashed](4\lxx/9,3\lyy/4)--(5\lxx/9,\lyy/2);
\draw[black,very thick,dashed](7\lxx/9,3\lyy/4)--(5\lxx/9,\lyy/2);
\draw[black,very thick,dashed](\lxx/4,\lyy/4)--(5\lxx/6,\lyy/4);
\draw[black,very thick,dashed](\lxx/2,\lyy/4)--(5\lxx/9,\lyy/2);
\draw[black,very thick,dotted](\lxx/4,\lyy/4)--(\lxx/4,9\lyy/5) node [pos=1,above=2] {\textcolor{black}{$z=\bar{z}=0$}};
\end{tikzpicture}
        \caption{Poincar\'e coordinates}
        \label{fig:A}
    \end{subfigure}
    \begin{subfigure}[b]{0.48\linewidth}        %% or \columnwidth
        \centering
       \includegraphics[width = .6\textwidth]{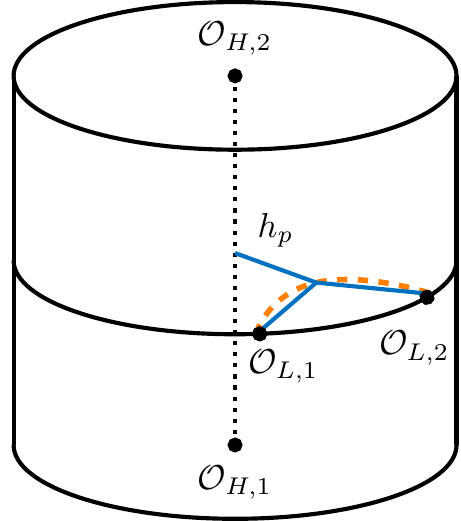}
        \caption{Global coordinates}
        \label{fig:B}
    \end{subfigure}
    \caption{Bulk setup for computing a heavy-light semiclassical Virasoro block.   The heavy operators $\Oc_{H_{1,2}}$ set up a conical defect geometry centered at the dotted line in the bulk.  The conical defect sources a bulk field dual to the exchanged primary operator $\Oc_p$.  The external light operators $\Oc_{L_{1,2}}$ interact with the bulk field along a geodesic; in particular, the interaction vertex is to be integrated over the bulk geodesic (dashed orange line) connecting the light operator insertion points. In the Poincar\'e figure, the corresponding Virasoro block in the CFT is indicated by the dashed black lines.}\label{fig:vir}
\end{figure}
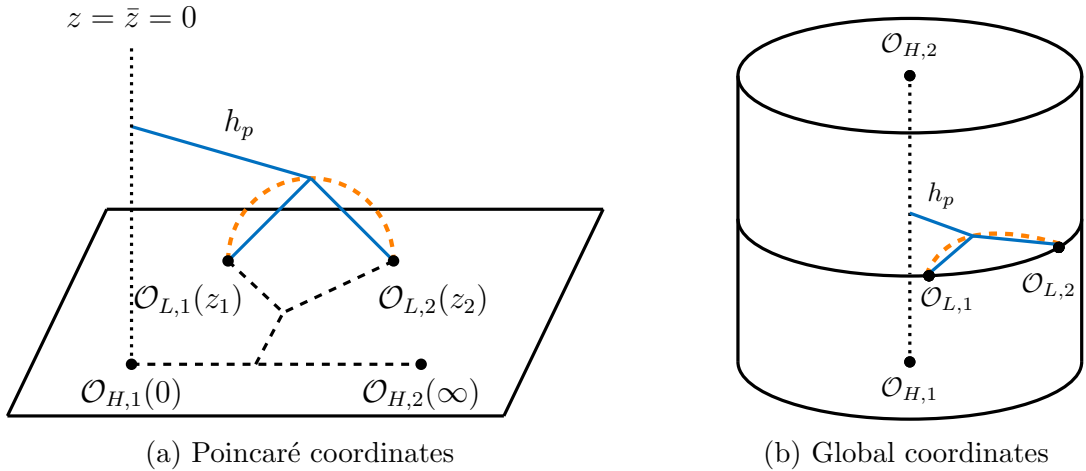

Now we need to generalize this to $\alpha \neq 1$. This is equivalent to taking the heavy-light limit \eqr{hllim} instead of keeping all $h_i$ fixed. In the geodesic Witten diagram picture, we now want to ``scale up'' the dimensions $h_1$ and $h_2$ with large $c$. This suggests a rather natural proposal: let $\g_{12}$ backreact, and evaluate (\ref{geoblock}) in the new spacetime. This is most naturally phrased if, as in (\ref{FKWw}), we take the heavy operators to be located at past and future infinity. Then the $\g_{12}$ geodesic, which sits at $\rho=0$, will backreact on the AdS$_3$ geometry to generate a conical defect or BTZ black hole, with metric \eqr{condef}. To obtain the heavy-light Virasoro block, we still compute (\ref{geoblock}), but now with the propagators for the light operators being defined in the conical defect metric (\ref{condef}) that is produced by the heavy operators.\footnote{Actually, in the next section this statement will be refined slightly.} We can think of this as the conical defect and light particle geodesic exchanging a bulk field corresponding to the primary $\Oc_p$. This provides a pleasingly intuitive picture for the heavy-light Virasoro block. We have drawn this setup in Figure \ref{fig:vir}, and in the middle frame of Figure \ref{f1a}.

We may also reason starting from the worldline picture described in Section \ref{review}, which computes the Virasoro block in the limit of fixed $h_i/c$ and $h_p/c$. We can obtain the heavy-light block by ``undoing'' the saddle-point approximation for the propagation of the light fields $h_{L_1,L_2,p}$, while keeping the conical defect geometry sourced by the heavy fields. This again suggests the picture in terms of the geodesic Witten diagram in the conical defect background.  Actually, at first glance there appears to be a mismatch between the worldline picture in \cite{Hijano:2015rla} and the approach presented here.  Namely, in \cite{Hijano:2015rla} the worldlines of the light fields meet at a vertex whose location is found by minimizing the total worldline action.  The location of this vertex typically does not lie on the geodesic connecting the external light operators.  By contrast, here the interactions are constrained to occur on the geodesic.  Despite this apparent difference, the results agree, as we explain in appendix \ref{worldapp}.

In the remainder of this section we verify our prescription by direct computation, showing how the bulk diagram reproduces (\ref{FKWw}). We will restrict to the case that all operators are spinless, obeying $h=\hb$ (in the next section we consider the case $h\neq 0$ and $\hb=0$.)
\subsection{Evaluating the geodesic Witten diagram}

We now reproduce \eqr{FKWw} using the geodesic Witten diagram in the conical defect background. Let us use CFT$_2$ notation to denote this as $\cW_{2h_p,0}$. With the operators at the specified configurations, we want to compute
\es{wdef}{{\cal W}_{2h_p,0}(w) \!\equiv\! \int_{-\infty}^\infty\! d\lambda\!\int_{-\infty}^\infty\!& d\lambda' G_{b\p}(\tau_1=-\infty,\tau(\l))G_{b\p}(\tau_2=\infty,\tau(\l) )\\
& \times G_{bb}^{(\a)}(y(\lambda),y(\lambda'); 2h_p) G_{b\p}^{(\a)}(w_1=0,y(\lambda'))G_{b\p}^{(\a)}(w_2=w,y(\lambda')) \text{ ,}}
where $G^{(\a)}$ is a propagator in the conical defect metric \eqr{condef}. In the first line, we have specifically highlighted the $\tau$-dependence to make clear that these operators generate in- and out-states on the cylinder. We will assemble this integrand piece-by-piece.

We first recall a few facts. A bulk scalar field $\phi$ dual to a CFT operator $\Oc$ of dimension $(h,h)$ has mass $m^2=4h(h-1)$ and obeys $(\nabla^2 -m^2)\phi=0$ in the absence of interactions. The bulk-to-boundary propagator in global AdS (i.e. $\alpha=1$) is
\be\label{bpprop}
G_{b\p}(x',y)= \left( \cos \rho \over \cosh(\tau-\tau')-\sin\rho\cos(\phi-\phi') \right)^{2h}  \text{ .}
\ee
Similarly, the scalar bulk-to-bulk propagator in global AdS, which obeys the wave equation with a delta function source, is,
\be\label{bbprop}
G_{bb}(y,y'; 2h)=\xi^{2h}{_2{F_1}}(h,h+{1\over 2},2h;\xi^2) = {e^{-2h\sigma(y,y')}\over e^{-2\sigma(y,y')}-1} \text{ ,}
\ee
where $\xi$ is related to the chordal distance $\xi^{-1}-1$, and $\sigma(y,y')$ is the geodesic distance between the two bulk points:
\be\label{bbprop2}
\sigma(y,y') = \ln\left(1+\sqrt{1-\xi^2}\over \xi\right)~,\quad \xi = { \cos \rho \cos \rho' \over \cosh(\tau-\tau')-\sin \rho \sin\rho' \cos(\phi-\phi')}~.
\ee

In \eqr{wdef}, we need to evaluate the propagators for the light external and internal operators in the conical defect geometry. We will obtain these by taking the global AdS results \eqr{bpprop}--\eqr{bbprop2} and making the replacements $\tau\rt \alpha \tau$ and $\phi\rt \alpha \phi$, which takes the metric to that of the conical defect. It should be noted that this does not in fact produce the proper bulk-to-boundary propagator for the conical defect, because the periodicity $\phi\cong \phi+2\pi$ is not respected. Therefore, the Virasoro block computed using this propagator will not be single-valued under $\phi\cong \phi+2\pi$. However, this is in fact what we want, because the Virasoro blocks have a branch cut and are not single-valued. This branch cut will be correctly reproduced using these non-single-valued propagators.

Having established that, we begin our calculation. The product of heavy operator propagators, with endpoints anchored at past and future infinity, and evaluated at $\rho=0$, is (dropping a prefactor)
\be\label{gheavy}
G_{b\p}(\tau_1=-\infty,\tau)G_{b\p}(\tau_2=\infty,\tau)= e^{-2H_{12}\tau} \text{ .}
\ee
Noting that $\lambda=\a\tau$ is proper time at the origin, \eqr{gheavy} gives the first line of \eqr{wdef}. For the light external and internal operators we use the bulk-to-boundary propagator in the conical defect geometry, as described above. To pull them back to the geodesics we need an expression for the geodesic itself, connecting the insertion points of the external light operators. Consider a geodesic beginning and ending at points $w_1$ and $w_2$ on the boundary, respectively. To simplify matters, we will take the two points to lie on a common time slice, so that $w_{12}=w_1-w_2$ is real. We then have
\be \cos\rho(\lambda) = {\sin{\alpha w_{12}\over 2}\over \cosh \lambda}~,\quad e^{2i\alpha w(\lambda)} = {\cosh(\lambda -{i\alpha w_{12}\over 2})\over \cosh(\lambda +{i\alpha w_{12}\over 2})}e^{i\alpha (w_1+w_2)} \text{ .}
\ee
The bulk-to-boundary propagators for the light fields evaluated on the geodesic then work out to be
\be
G^{(\a)}_{b\p}(w_1,y(\lambda'))= {e^{-2h_{L_1}\lambda'}\over (\sin{\alpha w_{12} \over 2})^{2h_{L_1}} }~,\quad G^{(\a)}_{b\p}(w_2,y(\lambda'))= {e^{2h_{L_2}\lambda'}\over (\sin{\alpha w_{12} \over 2})^{2h_{L_2}} } \text{ .}
\ee
Plugging into \eqr{wdef}, we set $w_1=0, w_2=w$. Finally, the bulk-to-bulk propagator for the field of dimension $h_p$ evaluated with one endpoint at $\rho=0$ at time $\tau$, and the other on the geodesic at time $\tau'=0$, is
\be
G^{(\a)}_{bb}(y(\lambda),y(\lambda'); 2h_p)=\xi^{2h_p}{_2{F_1}}(h_p,h_p+{1\over 2},2h_p;\xi^2)~,\quad \xi={\sin{\alpha w_{12}\over 2} \over \cosh \lambda \cosh \lambda'} \text{ .}
\ee
Putting everything together, we get the following integral expression
\es{result}{
\cW_{2h_p,0}(w)&= \lr{\sin\tfrac{\alpha w}2}^{2h_p-2h_{L_1}-2h_{L_2}} \int_{-\infty}^\infty\! d\lambda\int_{-\infty}^\infty\! d\lambda' e^{-{2H_{12} \over \alpha}\lambda-2h_{12}\lambda'}(\cosh \lambda\cosh \lambda')^{-2h_p}\\
&\quad\quad\quad\quad\quad\quad\quad\quad\quad\quad\quad\quad\times {_2{F_1}}\left(h_p,h_p+{1\over 2},2h_p;{(\sin{\alpha w\over 2})^2 \over (\cosh \lambda \cosh \lambda')^2}\right) \text{ .}}
The integrals can be evaluated by writing the series expansion of the hypergeometric function and using some identities. This is carried out in appendix \ref{integral} and the result is
\es{main}{\cW_{2h_p,0}(w) \propto\lr{\sin\tfrac{\alpha w}2}^{2h_p-2h_{L_1}-2h_{L_2}}\times\,&{_2{F_1}}\left(h_p+h_{12},h_p-{H_{12}\over \alpha} ,2h_p;1-e^{i\alpha w}\right)\\\times\,&{_2{F_1}}\left(h_p+h_{12},h_p-{H_{12}\over \alpha} ,2h_p;1-e^{-i\alpha w} \right)~,}
which matches \eqr{FKWw}. (Recall that we have systematically dropped all normalization factors.) This is one of our main results.
\vs\vs
It is also illuminating to reduce the expression for $\cW_{2h_p,0}(w)$ to a single integral as follows. Consider the part of the integral depending on $\lambda$,\footnote{This field solution was denoted $\varphi^{12}_{\D}(y')$ in \cite{Hijano:2015zsa}.}
\bea
\vphi_p(y')&=& \int_{-\infty}^\infty\! d\lambda G_{b\p}(\tau_1=-\infty,{\lambda\over \alpha})G_{b\p}(\tau_2=\infty,{\lambda\over \alpha} ) G_{bb}(y(\lambda),y';2h_p)\cr
&=&\alpha \int_{-\infty}^\infty\!d\tau e^{-2H_{12}\tau} G_{bb}(\rho=0,\tau; y'; 2h_p)  \text{ .}
\eea
In \eqr{wdef}, $y'$ is pulled back to the light geodesic, but we leave it general here. $\vphi_p(y')$ obeys $(\nabla^2-4h_p(h_p-1))\vphi_p=0$ away from a delta function source at $\rho=0$; is rotationally invariant; has a time dependence $e^{-2H_{12}\tau}$; and has normalizable falloff at the AdS boundary. These properties uniquely fix $\vphi_p$, and by solving the field equation in the conical defect background we find
\be\label{phisol}
\vphi_p(y') = (\cos \rho')^{2h_p} {_2{F_1}}\Big(h_p+{H_{12}\over \alpha},h_p-{H_{12}\over \alpha},2h_p; \cos^2\rho'\Big)e^{-2H_{12}\tau'} \text{ .}
\ee
The geodesic corresponding to the external light operators thus propagates in the conical defect dressed by the scalar field solution $\vphi_p(y(\l'))$ corresponding to the primary $\Oc_p$:
\es{}{\cW_{2h_p,0}(w)  &= \int_{-\infty}^\infty\! d\lambda' \,\vphi_p(y(\lambda') )\,G_{b\p}(w_1=0,y(\lambda'))\,G_{b\p}(w_2=w,y(\lambda')) \\
&= \left(\sin{\alpha w\over 2}\right)^{2h_p-2h_{L_1}-2h_{L_2} } \int_{-\infty}^\infty\! d\lambda'e^{-2h_{12}\lambda'-2{H_{12}\over \alpha}\l'} (\cosh\lambda')^{-2h_p} \\&\quad\quad\quad\quad\quad\quad\quad\quad\quad\quad\quad\quad\times{_2{F_1}}\left(h_p+{H_{12}\over \alpha},h_p-{H_{12}\over \alpha},2h_p; {\sin^2 {\alpha w\over 2}\over \cosh^2\lambda'} \right) \text{ .}}
This formula can also be seen to reproduce (\ref{FKWw}).
\vs\vs
To summarize the results of this section, we verified a simple bulk prescription for reproducing the semiclassical heavy-light Virasoro block, involving a light particle geodesic interacting with a heavy particle worldline via the exhange of a light intermediate field. To be precise, our computation doesn't quite allow us to extract the individual factors $\F$ and $\overline{\F}$ in (\ref{FKWw}) because of our restriction to real $w$. This limitation will be overcome in the next section.

%%%%%%%%%%%%%%%%%%%%%%%%%%%%%%%%%%%%%%%%%%%%%%%%%%%%%%%%%%%%%%%%%%%%%%%%%%%%%%%%
\section{Semiclassical Virasoro blocks from AdS$_3$ higher spin gravity}
\label{higherspin}
%%%%%%%%%%%%%%%%%%%%%%%%%%%%%%%%%%%%%%%%%%%%%%%%%%%%%%%%%%%%%%%%%%%%%%%%%%%%%%%%

In the previous section we used bulk scalar fields to compute semiclassical Virasoro blocks. However, since bulk scalar fields are dual to CFT operators with equal holomorphic and anti-holomorphic dimensions, this computation in fact gave the product of the holomorphic and anti-holomorphic Virasoro blocks. It is interesting to ask whether there is a bulk computation that yields the holomorphic block, say, directly. In this section we provide such a computation.

To proceed, we take all operators to have vanishing anti-holomorphic dimension. If the holomorphic dimensions are restricted to be positive integers $s$, then a description in terms of dual bulk fields is available. Namely, such a spin-$s$ conserved current in the CFT is dual to a massless spin-$s$ field in the bulk, the latter being described by a symmetric transverse traceless tensor of rank-$s$. Like the graviton in three dimensions, such fields have no local degrees of freedom. Working in terms of these fields, we will show how to extract the holomorphic Virasoro block. Our computation will only directly yield the block for operators of integer dimension; however, this is not really a limitation once we use a known property of the Virasoro blocks. Namely, after stripping off a prefactor, the block admits a series expansion in $z$ whose coefficients are rational functions of the operator dimensions. Knowing the rational functions for integer values of the dimensions is clearly sufficient to determine the functions in general.

Rather than working in terms of symmetric traceless tensors, it will be extremely convenient to use an equivalent Chern-Simons formulation \cite{Witten:1988hc,Achucarro:1987vz,Campoleoni:2010zq}. The reason is that in the Chern-Simons formulation the gauge algebra and action separate into two parts, corresponding to the holomorphic factorization of the chiral algebra. To extract the holomorphic part we need only deal with a single half of the Chern-Simons theory. This factorization is much less obvious in the tensor formulation. As we will see, our entire computation will reduce to performing various gauge transformations

In the following, we begin with a brief review of the needed aspects of higher spin gravity in the Chern-Simons formulation. We then describe the setup of our computation, including its subtleties, and finally present the details.

\bigskip

\subsection{Higher spin fields in AdS$_3$}\label{subsec:hs}

\subsubsection{Review of Chern-Simons description}

 Relevant background can be found in \cite{Campoleoni:2010zq,Campoleoni:2011hg,Gaberdiel:2011wb,Ammon:2012wc}. Three dimensional general relativity is equivalent to a Chern Simons gauge theory with gauge group $G=SL(2,\mathbb{R})\times SL(2,\mathbb{R})/\mathbb{Z}_2$ \cite{Achucarro:1987vz,Witten:1988hc}. To see this, we start from the dreibein $e_{\mu}^{a}$ and the spin connection $\omega_{\mu}^{a}$ of the first order formulation of gravity. These objects can be combined into a pair of Chern Simons connections, each valued in a different copy of the algebra $sl(2,\mathbb{R})$,
\begin{equation}\label{eq:connections}
A=L_m \left( \omega_{\mu}^{m}+{{1}\over{l}}e_{\mu}^{m} \right)dx^\mu\text{ ,} \quad\quad \bar{A}=L_m \left( \omega_{\mu}^{m}-{{1}\over{l}}e_{\mu}^{m} \right)dx^\mu \text{ ,}
\end{equation}
where $L_m$ with $m=\{-1,0,1\}$ are a convenient choice of generators of the $sl(2,\mathbb{R})$ algebra obeying $[L_m,L_n]=(m-n)L_{m+n}$. Up to a total derivative, the Einstein-Hilbert action can be written in terms of these connections as the difference of two Chern Simons actions with level $k=l/4G_3$, $l$ being the AdS$_3$ radius and $G_3$ being the three-dimensional Newton constant,
\begin{equation}\label{eq:CSaction}
I_{EH}=I_{CS}[A]-I_{CS}[\bar{A}]\text{ ,}\quad \text{with }\quad I_{CS}[A]={{k}\over{4\pi}}\int \text{tr}\left( A\wedge dA+{{2}\over{3}}A\wedge A\wedge A \right) \text{ .}
\end{equation}
Here the trace $\text{tr}$ stands for the symmetric bilinear form on $sl(2,\mathbb{R})$. The Euler-Lagrange equations imply that $A$ and $\bar{A}$ are flat connections, and indeed these are equivalent to Einstein's equations under the dictionary (\ref{eq:connections}). The metric can be recovered from the connections via
\begin{equation}
g_{\mu_1 \mu_2}=\text{tr}\left( e_{\mu_1}e_{\mu_2 } \right)\text{ ,}
\end{equation}

Another salient aspect is the relation between gauge symmetries in the Chern-Simons and metric formulations. In the latter, we have local translations ($\xi$) and local Lorentz transformations ($\lambda$), under which the the dreibein and spin connection transform as
\begin{equation}
\begin{aligned}
\delta e &= d\xi+[\omega,\xi]+[e,\lambda] \text{ ,}\\
\delta \omega &=d\lambda+[\omega,\lambda]+{{1}\over{l^2}}[e,\xi] \text{ .}
\end{aligned}
\end{equation}
These are related to $sl(2,\mathbb{R})$-valued gauge parameters $\Lambda$ and $\bar{\Lambda}$ that transform the connections as follows
\begin{equation}
\begin{aligned}
\Lambda&={{1}\over{l}}\left( \xi + \lambda \right)\quad \text{and} \quad \delta A=d A +[A,\Lambda] \text{ ,}\\
\bar{\Lambda}&={{1}\over{l}}\left( \xi - \lambda \right)\quad \text{and} \quad \delta \bar{A}=d \bar{A} +[\bar{A},\bar{\Lambda}] \text{ .}
\end{aligned}
\end{equation}

To generalize this theory to include higher spin fields, one enlarges the gauge algebra to some ${\cal G}\supset sl(2,\mathbb{R}) \times sl(2,\mathbb{R})$. The decomposition of the adjoint of $\cG$ into $sl(2,\mathbb{R})$ representations determines the spectrum of higher spin fields around the AdS$_3$ vacuum. It will be sufficient for our purposes to consider the simple case $\cG=sl(N,\mathbb{R})\times sl(N,\mathbb{R})$ for some integer $N>2$. With a principally embedded $sl(2,\mathbb{R}) \times sl(2,\mathbb{R})$ subalgebra, this describes AdS$_3$ gravity coupled to additional non-propagating massless fields of spin $3,4, \ldots N$. The formulas above for the Chern-Simons action and gauge transformations still apply, just with $sl(2,\mathbb{R})$ replaced by $sl(N,\mathbb{R})$. The symmetric traceless spin-$s$ tensor of the metric formulation may be obtained as an order-$s$ polynomial in the generalized $sl(N,\mathbb{R})$ dreibein. For $s=3$, for example,
\begin{equation}\label{tensors}
\psi^{(3)}_{\mu_1 \mu_2 \mu_3}\propto \text{tr}\left( e_{(\mu_1}e_{\mu_2 }e_{\mu_3 )} \right) \text{ .}
\end{equation}
In Appendix \ref{subsec:slnRalgebra} we set our conventions for the $sl(N,\mathbb{R})$ algebra. We write the generators as $W^{(s)}_m$ with $m=-(s-1), \ldots s-1$ and $s= 2, 3, \ldots N$.
The $sl(2,\mathbb{R})$ subalgebra is generated by $L_m\equiv W^{(2)}_m$. The set of generators of fixed $s$ fill out a $2s-1$ dimensional irreducible representation of $sl(2,\mathbb{R})$.

Let us now say more about the form of the connections we will be using. Henceforth, we refer only to the connection $A$, as $\overline{A}$ will play no role in our computations. In Poincar\'e coordinates, AdS$_3$ has metric
\e{}{ds^2 = {du^2 + dzd\zb\over u^2} \text{ .}}
It is useful to choose a gauge where the connections adopt a simple form that permits easy comparison with CFT. For this we first introduce the radial gauge, in which the connections read
\begin{equation}
A= b^{-1}d b+b^{-1}a b~,\quad b=u^{-L_0} \text{ ,}
\end{equation}
where $a$ is a flat $sl(N,\mathbb{R})$ valued one-form of the form
\e{}{a=a_z(z,\zb)dz+a_{\zb}(z,\zb)d\zb~.}
Poincar\'e AdS$_3$ corresponds to the choice $a=L_1 dz$. An asymptotically AdS$_3$ connection can be written in so-called highest weight gauge as
\begin{equation}\label{eq:hw}
a=\left(L_1+\sum^{N}_{s=2} J^{(s)}(z) W^{(s)}_{-(s-1)}\right)dz \text{ .}
\end{equation}
Flatness forces $\partial_{\zb}J^{(s)}=0$.
The asymptotic symmetry algebra is obtained by finding the most general gauge transformation that preserves the form (\ref{eq:hw}). Expanding in modes, one thereby arrives at the classical ${\cal W}_N$ algebra \cite{cfpt, gabhart, hr}, and $J^{(s)}(z)$ is identified with the vev of a spin-$s$ conserved current in the boundary theory.

\subsubsection{Correlation functions}

We now discuss the computation of correlation functions (e.g. \cite{Ammon:2011ua,Hijano:2013fja}).
The two-point function of currents can be defined as the response to an infinitesimal source coupled to the current. A delta function source for a spin-$s$ current is described by
\be
a_{\zb} = \mu^{(s)} \delta^{(2)}(z-z_1)W^s_{s-1} + \ldots \text{ ,}
\ee
where the $\ldots$ denote terms proportional to generators with lower mode index $m$, that will be induced by the flatness condition. To linear order in $\mu^{(s)}$, flatness (along with the highest weight gauge condition) implies that $a_z$ takes the form in (\ref{eq:hw}) with
\be\label{Js}
J^{(s)}(z_2) \propto {\mu^{(s)} \over (z_2-z_1)^{2s} }~.
\ee
The current-current two-point function is thus $\langle J^{(s)}(z_2)J^{(s)}(z_1)\rangle= {1\over (z_2-z_1)^{2s}}$, which is of course the result dictated by conformal invariance.

More generally, we can compute $n$-point functions of conserved currents. The idea is to demand that the currents have prescribed singularities as above at $n-1$ points,
\be
J^{(s_i)}(z) \sim {\mu^{(s_i)} \over (z-z_i)^{2s_i} } \quad {\rm as}\quad z\rightarrow z_i~,\quad i=1, 2, \ldots n-1  \text{ .}
\ee
We then impose flatness and compute $J^{(s_n)}(z_n)$. The term proportional to $\mu^{(s_1)}\mu^{(s_2)}\ldots \mu^{(s_{n-1})}$ is identified as $\langle J^{(s_1)}(z_1) \ldots J^{(s_n)}(z_n) \rangle$.

The above rules will yield the correct correlation functions on the plane because it is easy to see that gauge invariance implies that they will obey the correct Ward identities \cite{Gutperle:2011kf,deBoer:2013gz,deBoer:2014fra}, and these are sufficient to fix the result.  Equivalently, the correlator is a meromorphic function with singularities fixed by the OPE, which is again enough information to uniquely determine the answer.  Nonetheless, there is something a bit odd about our procedure.  Returning to the case of the two-point function, the form of the current in (\ref{Js}) translates into the following form for $A_z$,
\be\label{Aform}
A_z = {1\over u}L_1 + {\mu^{(s)} u^{s-1}\over (z-z_1)^{2s}}W^{(s)}_{-(s-1)}~.
\ee
From this we can work out the corresponding component of the symmetric traceless rank-$s$ tensor $\psi^{(s)}_{z\ldots z}$, using (\ref{tensors}).   We would then like to identify this result with the corresponding component of the bulk-to-boundary propagator for this field.  However, it can hardly escape notice that (\ref{Aform}) is singular at $z=z_1$ for all $u$, whereas standard bulk-to-boundary propagators are nonsingular in the interior of AdS, instead having a prescribed delta function in their near-boundary expansion. The simplest example that makes this distinction clear is the case of a $U(1)$ Chern-Simons gauge field.  In our construction we effectively use the bulk-to-boundary propagator $G_{zz}(u,z;z_1)\propto {1\over (z-z_1)^2}$, whereas a ``standard" computation, e.g. \cite{Dong:2014tsa}, of the bulk-to-boundary propagator yields
\be\label{dongprop}
G_{zz}(u,z;z_1) \propto \p_z \left( \zb -\zb_1 \over u^2 + |z-z_1|^2 \right)~.
\ee
This result agrees with the previous one as $u\rightarrow 0$, but is nonsingular.  The two results are clearly related by a singular gauge transformation that goes to zero at the boundary.  Due to this last property, we expect the two versions to give the same results for boundary correlators.  Given this, it is more convenient to use the version which has no dependence on $\zb$, which is what we are doing in using (\ref{Aform}).

\subsection{Setup}\label{subsec:setuphs}

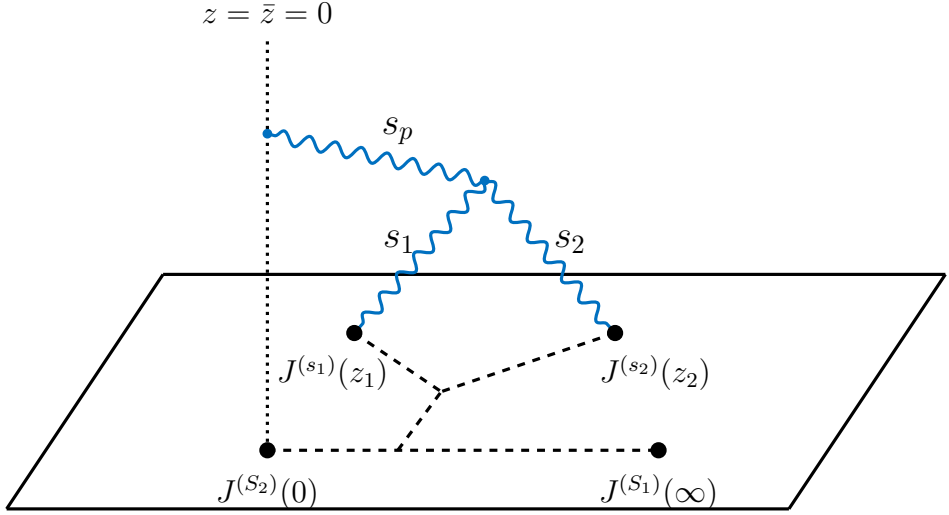
\begin{figure}
\begin{center}
\begin{tikzpicture}[scale=0.52]
\draw[black,very thick] (0,0) -- (\lx,0);
\draw[black,very thick] (0,0) -- (\lx/5,\ly);
\draw[black,very thick] (\lx,0) -- (\lx+\lx/5,\ly);
\draw[black,very thick] (\lx/5,\ly) -- (6\lx/5,\ly);

\draw[black,very thick,fill=black](\lx/3,\ly/4) circle (\lx/120);
\draw (\lx/3,\ly/4) node[below=5] {$J^{(S_2)} (0)$};

\draw[black,very thick,fill=black](5\lx/6,\ly/4) circle (\lx/120);
\draw (5\lx/6,\ly/4) node[below=5] {$J^{(S_1)}  (\infty)$};

\draw[propagatorblue,very thick,snake it] (5.5\lx/9,7\ly/5) -- (4\lx/9,3\ly/4) node [pos=0.4,left=2] {\textcolor{black}{\large $s_1$}};
\draw[propagatorblue,very thick,snake it] (5.5\lx/9,7\ly/5) -- (7\lx/9,3\ly/4) node [pos=0.4,right=2] {\textcolor{black}{\large $s_2$}};
\draw[propagatorblue,very thick,snake it] (5.5\lx/9,7\ly/5) -- (\lx/3,8\ly/5) node [pos=0.4,above=2] {\textcolor{black}{\large $s_p$}};

\draw[black,very thick,fill=black](4\lx/9,3\ly/4) circle (\lx/120);
\draw (4\lx/9,3\ly/4) node[below=15,left=-17] {$J^{(s_1)}  (z_1)$};

\draw[black,very thick,fill=black](7\lx/9,3\ly/4) circle (\lx/120);
\draw (7\lx/9,3\ly/4) node[below=15,right=-10] {$J^{(s_2)} (z_2)$};

\draw[black,very thick,dashed](4\lx/9,3\ly/4)--(5\lx/9,\ly/2);

\draw[black,very thick,dashed](7\lx/9,3\ly/4)--(5\lx/9,\ly/2);

\draw[black,very thick,dashed](\lx/3,\ly/4)--(5\lx/6,\ly/4);

\draw[black,very thick,dashed](\lx/2,\ly/4)--(5\lx/9,\ly/2);

\draw[black,very thick,dotted](\lx/3,\ly/4)--(\lx/3,2\ly) node [pos=1,above=2] {\textcolor{black}{ $z=\bar{z}=0$}};

\draw[propagatorblue,very thick,fill=propagatorblue](\lx/3,8\ly/5) circle (\lx/240);

\draw[propagatorblue,very thick,fill=propagatorblue](5.5\lx/9,7\ly/5) circle (\lx/240);
\end{tikzpicture}
\end{center}
\caption{Setup for computing the chiral heavy-light semiclassical Virasoro block using higher spin gauge fields, analogous to our previous construction using of scalar fields. Each gauge field is dual to a higher spin current. The heavy operators in this picture are $J^{(S_1)}$ and $J^{(S_2)}$, whose spins are taken to infinity. Rather then computing the diagram using propagators and vertices, we will obtain it through the use of higher spin gauge transformations, taking advantage of the fact that massless fields of positive integer spin in three-dimensions have no local degrees of freedom. }
\label{fig:hs}
\end{figure}

We aim to compute the semiclassical conformal block
\be
\langle J^{(S_1)}(\infty)J^{(S_2)}(0) P_{s_p}J^{(s_1)}(z) J^{(s_2)}(1)\rangle  \text{ .}
\ee
$J^{(S_{1,2})}$ correspond to heavy operators, in the same sense as in the previous section. Let us define
\e{}{s_{12} \equiv s_1-s_2~, \quad S_{12} \equiv S_1-S_2}
Our basic strategy is the same as before: the heavy operators set up a background solution, which we then dress with the field of dimension $s_p$. The fields of dimension $s_{1,2}$ propagate in this background. See Figure \ref{fig:hs}. The conformal block will then correspond to the two-point function for the $s_{1,2}$ fields evaluated in this background.

First consider the bulk description of the heavy operators. A geodesic at the origin of global coordinates maps to the radial geodesic $z=\zb=0$ in Poincar\'{e} coordinates. The expectation value of the stress tensor in the state dual to an operator of dimension $(S,0)$ is
\be
{\langle S |T(z) |S\rangle \over \langle S | |S\rangle}= {\langle \Oc_S(\infty) T(z) \Oc_S(0)\rangle \over \langle \Oc_S(\infty) \Oc_S(0)\rangle} = {S\over z^2}~.
\ee
 The connection describing the heavy particle geodesic is therefore
\be
a = \left(L_1 + {S\over z^2}L_{-1}\right)dz \text{ .}
\ee
It will be convenient to note that the stress tensor can be induced by a conformal transformation, which in turn can be described by a gauge transformation acting on the connection. Using the Schwarzian derivative transformation law for the stress tensor, we have that the above stress tensor is induced by the conformal transformation $z \rightarrow z' = z^\alpha$ with $\alpha=\sqrt{1-{24\over c}S}$. What we will do is to first take $S=0$ (or rather, set to zero the part of $S$ that scales like $c$), and then restore it at the end by applying this conformal transformation.

With this in mind, our starting point is the connection for Poincar\'{e} AdS with the addition of a delta function source at $z_2=1$ for a spin-$s_2$ field. We display only $a_z$, keeping in mind that various delta function terms in $a_{\zb}$ are implied by flatness,
\begin{equation}\label{eq:connections1}
a=\left( L_1 + J^{(s_2)}(z) W^{(s_2)}_{-(s_2-1)}\right) dz \text{ ,}
\end{equation}
with
\begin{equation}\label{eq:phis1}
J^{(s_2)}(z)=\frac{\mu_2}{\left(z-z_{2}\right) ^{2 s_{2}}} \text{ .}
\end{equation}
Justification for this was given in the previous subsection.

The next step, which is where all the work lies, is to perform a gauge transformation that turns on the exchanged field with spin $s_p$. This field is sourced at the heavy particle geodesic. The effect of this gauge transformation will be to induce a spin $s_1$ current $J^{(s_1)}(z)$, and from this we read off the conformal block.

Before diving into the details, let us discuss one aspect of this computation that deserves to be understood better.   In the previous section in which we computed the semiclassical Virasoro block using scalar fields, an important part of the story was that the external light operators only couple to the exchanged bulk field along a geodesic.  As discussed at length in \cite{Hijano:2015zsa} in the context of global blocks,  if the interaction vertex is integrated over all of AdS instead then the result is not a single conformal block, but rather an infinite sum of conformal blocks that includes the exchange of double-trace operators built out of the light operators. This is the case for ordinary Witten diagrams. In our higher-spin computation we use gauge transformations rather than integrating a vertex location, and so it is not obvious how to incorporate the different alternatives for how the vertex location is to be integrated.  Our procedure, which essentially amounts to computing the two-point function in the presence of a heavy background dressed with a spin-$s_p$ field, turns out to compute a single conformal block: no double trace exchanges appear.   It is convenient that the simplest prescription generates a single conformal block; however, it would also be useful to know how to modify the prescription so as to incorporate the light double-trace exchanges one expects from a full Witten diagram.  Perhaps this is related to our discussion in the previous subsection about choice of propagators.

\subsection{Detailed calculation}\label{subsec:calchs}

As described above, our starting point is the $sl(N,\mathbb{R})$ connection (\ref{eq:connections1}). We denote by $\Lambda_p$ the subsequent gauge transformation that turns on the spin $s_p$-field sourced at the geodesic. We demand that it obey
\be\label{eq:EqDeltaA}
\delta_{\Lambda_p} a = \left(J^{(s_p)} W^{(s_p)}_{-(s_p-1)}\right)dz~,\quad J^{(s_p)} \sim q_p z^{S_{12}-s_p}\quad {\rm as}\quad z \rightarrow 0  \text{ .}
\ee
The factor of $z^{S_{12}}$ is the Poincar\'{e} coordinate version of the time dependence in global coordinates employed in the last section.  That is, in (\ref{phisol}) we had $\varphi_p \sim e^{-2H_{12}\tau}$.  Writing $H_{12}=S_{12}$,  $z=e^{iw}$ with $w=\phi+i\tau$, this becomes $\varphi_p \sim (z\zb)^{S_{12}}$, and for a gauge field, as opposed to a scalar, only the holomorphic part appears.  The factor of $z^{-s_p}$ represents a constant current in global coordinates, transformed to Poincar\'{e} coordinates:  $J^{(s_p)}(z) = \left(dz\over dw\right)^{-s_p}J^{(s_p)}(w) \propto z^{-s_p}$.

We wish to solve this problem to order $\mu_2 q_p$. Furthermore, we demand that $\Lambda_p(z_2)=0$ since a nonzero gauge transformation at $z_2$ would mean that $\mu_2$ is no longer the coefficient of the source for the spin $s_2$ field. To implement the perturbation theory we write
\begin{eqnarray}
\Lambda_{p} &=&q_p (\Lambda^{\left( 0\right) }+ \mu_2 \Lambda^{\left( 1\right) }) , \\
a &=&a^{(0)}+\mu_2 a^{(1)} \text{ ,} \\
\delta a &=& q_p (\delta a^{(0)} + \mu_2 \delta a^{(1)})  \text{ ,}
\end{eqnarray}
with, from \eqref{eq:connections1},
\begin{equation}
a^{(0)}=L_{1}dz\text{ \ \ and\ \ \ }a^{(1)}={{1}\over{(z-z_2)^{2s_2}}} W_{-(s_{2}-1)}^{(s_{2})}dz \text{ .}
\end{equation}

We first work out $\Lambda^{\left( 0\right) }$. In this
case only spin $s_{p}$ generators are needed. We write the gauge
transformation as a linear combination of all algebra generators of spin $s_{p}$
\begin{equation}
\Lambda^{(0)}=\sum\limits_{m=-(s_{p}-1)}^{s_{p}-1}y_{m}^{(s_{p})}W_{m}^{(s_{p})} \text{ ,}
\end{equation}
under which  the connection transforms as follows:
\begin{equation}
\begin{aligned}\label{eq:da0}
\delta a^{(0)} &=d\Lambda ^{(0)}+\left[ a^{(0)},\Lambda ^{(0)}\right] \\
&=\sum\limits_{m=-(s_{p}-1)}^{s_{p}-1}\left\{ \partial
_{z}y_{m}^{(s_{p})}W_{m}^{(s_{p})}+\left[ \left( s_{p}-1\right) -m\right]
y_{m}^{(s_{p})}W_{m+1}^{(s_{p})}\right\} dz \text{ .}
\end{aligned}
\end{equation}
where we used $[L_1,W^{(s_p)}_m] = [(s_p-1)-m]W^{(s_p)}_{m+1}$. Equation (\ref{eq:EqDeltaA}) is now equivalent to the following set of coupled
differential equations:
\begin{equation}
\partial _{z}y_{s_{p}-q+1}^{(s_{p})}+(q-1)y_{s_{p}-q}^{(s_{p})}=z^{S_{12}-s_p}\delta _{q,2s_{p}} \text{,}
\end{equation}
for $q\in \left[ 2,2s_p\right] $ and we are defining $y_{s}^{(s)}=y_{-s}^{(s)}=0$. One can solve this\ system by induction, and
the answer for the first $2s_p-2$ equations reads
\begin{equation}\label{eq:RecYm}
y_{s_{p}-q}^{(s_p)}=\frac{\left( -1\right) ^{q+1}}{\Gamma \left( q\right) }\partial _{z}^{q-1}y_{s_{p}-1}^{(s_{p})} \text{ .}
\end{equation}
We are left with a single differential equation for $y_{s_p-1}^{(s_p)}$. It
reads
\begin{equation}
\frac{1}{\Gamma \left( 2s_p-1\right) }\partial
_{z}^{2s_p-1}y_{s_{p}-1}^{(s_{p})}=z^{S_{12}-s_p} \text{ ,}
\end{equation}
and the solution obeying $\Lambda^{(0)}(1)=0$ is a hypergeometric function
\begin{equation}\label{eq:Ysm1}
y_{s_{p}-1}^{(s_{p})}=\left( 1-z\right) ^{2s_p-1} {}_2F_1(1,s_p-S_{12},2s_p; 1-z) \text{ .}
\end{equation}
With this we have fully obtained the gauge transformation $\Lambda ^{(0)}$.

We now turn to $\Lambda ^{(1)}$. We
can write this generally as a sum over all possible generators\footnote{We are assuming that $N\geq s_1, s_2, s_p$. These spins are kept arbitrary, so we are working at arbitrary $N$. The calculations therefore only require knowing the $sl(N,\mathbb{R})$ algebra. As will become clear in the following and in Appendix \ref{ap:Hsap}, the $N$-dependence of structure constants only affects overall normalization of the conformal blocks we are computing, and hence is moot for our purposes.}
\begin{equation}
\Lambda ^{(1)}=\sum\limits_{\tilde{s}=2}^{N }\sum\limits_{m=-(\tilde{s}-1)}^{\tilde{s}-1}\gamma _{m}^{(\tilde{s})}W_{m}^{(\tilde{s})} \text{ ,}
\end{equation}
and then
\begin{equation}\label{eq:da1}
\begin{aligned}
\delta a^{(1)} &=d\Lambda ^{(1)}+\left[ a^{(0)},\Lambda ^{(1)}\right] +\left[ a^{(1)},\Lambda ^{(0)}\right] \\
&=\sum\limits_{\tilde{s}=2}^{N }\sum\limits_{m=-(\tilde{s}-1)}^{\tilde{s}-1}\left\{ \partial _{z}\gamma _{m}^{(\tilde{s})}W_{m}^{(\tilde{s})}+\left[
\left( s_{f}-1\right) -m\right] \gamma _{m}^{(\tilde{s})}W_{m+1}^{(\tilde{s})}\right\} dz \\
&+\sum\limits_{m=-(s_p-1)}^{s_p-1}{{y_{m}^{(s_{p})}}\over{(z-z_2)^{s_2}}}
\sum\limits_{u=2,4,6...}^{s_{2}+s_p-\left\vert
s_{2}-s_p\right\vert -1}g_{u}^{s_{2}s_{p}}\left( - s_{2}+1
,m;N\right) W_{m-\left( s_{2}-1\right) }^{(s_{p}+s_{2}-u)}dz \text{ ,}
\end{aligned}
\end{equation}
where $g_{u}^{s_{2}s_{p}}$ are the structure constants of the commutator between $W^{(s_2)}_{-(s_2-1)}$ and $W^{(s_p)}_{m}$. (See Appendix \ref{ap:Hsap} for their explicit form.) Examining the range of the $u$ sum in equation \ref{eq:da1}, we see that only spins between $\left\vert s_{2}-s_{p}\right\vert +1$ and $s_{p}+s_{2}-2$ contribute to $\Lambda ^{(1)}$. This also means that the interaction of the fields of spin $s_{p}$ and $s_{2}$ gives rise to conformal blocks of spin $\tilde{s}=s_{1}$ within
those bounds. In order to remain in the highest weight gauge we demand
\begin{equation}\label{eq:EqDeltaA2}
\delta a^{(1)}=\sum\limits_{s_{1}=\left\vert s_{2}-s_{p}\right\vert
+1}^{s_{p}+s_{2}-2}J^{\left( s_{1}\right) }\left( z\right)
W_{-(s_{1}-1)}^{(s_{1})}dz~\text{ ,}
\end{equation}
where $J^{\left( s_{1}\right) }\left( z\right) $ are unknown functions that
determine the conformal blocks we are trying to calculate. Before solving
this set of equations for the parameters $\gamma _{m}^{(s_{1})}$, we need to make a choice for $\gamma _{m}^{(s_{1})}$ with $m\geq s_{p}-s_{2}$. The
simplest and most natural choice is to set them to zero.  The meaning of these parameters when they are nonzero is not clear to us. Equations (\ref{eq:EqDeltaA2}) now have a unique solution
for the parameters of the gauge transformation and the functions $J^{\left( s_{1}\right) }\left( z\right) $. The $\gamma _{m}^{(s_{1})}$ parameters can
be obtained recursively. As a function of $y_{m}^{(s_{p})}$, they read
\begin{equation}\label{eq:gamma}
\begin{aligned}
\gamma _{s_{p}-s_{2}-q}^{(s_{1})}
&=\sum\limits_{i=0}^{q-1}\sum\limits_{j=1}^{q-i}g_{s_{p}-s_{12}}^{s_{2}s_{p}}\left( -\left( s_{2}-1\right) ,s_{p}+i+j-1-q;N\right)\frac{\Gamma \left(
s_{1}+s_{2}-s_{p}+q-i-j\right) }{\Gamma \left( s_{1}+s_{2}-s_{p}+p\right) }\\
&\times {{i+j-1} \choose {i}} \partial _{z}^{i}(z-z_2)^{-2s_2}\left( -1\right) ^{j}\partial
_{z}^{j-1}y_{s_{p}+i+j-1-q}^{(s_{p})} \text{ ,}
\end{aligned}
\end{equation}
 The function $J^{\left( s_{1}\right) }\left( z\right) $ can be written as a
function of the parameters $\gamma _{m}^{(s_{1})}$. The result, after using (\ref{eq:gamma}), reads\footnote{Here and below, $\sim$ means that we drop overall prefactors.}
\begin{equation}\begin{aligned}\label{eq:J}
J^{\left( s_{1}\right) }\left( z\right) &\sim
\sum\limits_{i=0}^{s_{p}+s_{12}-1}\sum\limits_{j=1}^{s_{p}+s_{12}-i}  \left( i+j \right)_{s_{p}-s_{12}-1} \frac{\Gamma \left( 2s_{1}-i-j\right) }{\Gamma \left( 2s_{1}\right) }{{i+j-1} \choose {i}} \\
&\times \partial _{z}^{i}(z-z_2)^{-2s_2} \left( -1\right) ^{j}\partial
_{z}^{j-1}y_{i+j-1-s_{12}}^{(s_{p})} \text{ .}
\end{aligned}
\end{equation}
Here we have replaced the relevant structure constants  by a simple expression obtained in appendix \ref{subsec:slnRalgebra}, namely
\begin{equation}\label{eq:simpleg}
g_{s_{p}-s_{12}}^{s_{2}s_{p}}\left( - s_{2}+1 ,i+j-1-s_{12};N\right)  \sim \left( i+j \right)_{s_{p}-s_{12}-1}
 \text{ ,}
\end{equation}
where we have ignored an overall constant that does not depend on $i$ or $j$.
After this, the sums over $i$ and $j$ can be performed analytically (see appendix \ref{ap:Hsap}) to
obtain
\begin{equation}\label{eq:Js1}
J^{\left( s_{1}\right) }\left( z\right) \sim \left( 1-z\right)
^{s_{p}-s_{2}-s_{1}}{}_2F_1( s_{p}-S_{12},s_{p}+s_{12},2s_{p};1-z) \text{ .}
\end{equation}

The final step is to perform a coordinate transformation to take this result
from Poincar\'e AdS$_{3}$ to the geometry generated by heavy operators with
conformal dimensions $S_{1}, S_2$. The transformation reads
\begin{equation}
\begin{aligned}
z&\rightarrow z^{\prime }\left( z\right) =z^{\alpha }\text{ \ \ with \ \ }\alpha =\sqrt{1-\frac{24}{c}S_{1,2}} \text{ ,}
\end{aligned}
\end{equation}
where $S_{1,2}$ stands for either of $S_1$ or $S_2$, the distinction being subleading in $1/c$. Before writing the answer for the holomorphic Virasoro block, the meaning of $S_{12}$ has to be reinterpreted in the new coordinates. Before the coordinate transformation, we had the following expression for the three-point function among the two heavy operators and the exchanged light operator.
\begin{equation}
\langle J^{(S_1)}(\infty) J^{(S_2)} (0) J^{(s_p)}(z) \rangle \sim z^{S_{12}-s_p} \text{ .}
\end{equation}
This is the expression used previously in (\ref{eq:EqDeltaA}) for the expectation value of the spin-$s_p$ conserved current in the presence of the heavy operators. After the coordinate transformation $z\rightarrow z^{\prime}(z)$ this correlator reads
\begin{equation}
\langle J^{(S_1)}(\infty) J^{(S_2)} (0) J^{(s_p)}(z') \rangle \sim \left(z^{\prime}\right)^{{{S_{12}}\over{\alpha}}-s_p} \text{ .}
\end{equation}
This shows that after performing the coordinate transformation, $S_{12}$ has to be adjusted to $S_{12}/\alpha$. The final answer for the Virasoro conformal block can finally be written down:
\begin{equation}
\begin{aligned}
\langle J^{(S_1)}(\infty)J^{(S_2)}(0) P_{s_p}J^{(s_1)}(z)J^{(s_2)}(1)\rangle &\sim \left( \frac{\partial
z^{\prime }}{\partial z}\right) ^{s_{1}}\left( \frac{\partial z^{\prime }}{\partial z}\right) ^{s_{2}} \left. J^{\left( s_{1}\right) }\left( z^{\alpha }\right) \right|_{S_{12} \rightarrow {{S_{12}}\over{\alpha}}}
\\
&\sim z^{s_{1}\left( \alpha -1\right) }\left( 1-z^{\alpha }\right)
^{s_{p}-s_{2}-s_{1}} {}_2F_1( s_{p}-{{S_{12}}\over{\alpha}},s_{p}+s_{12},2s_{p};1-z^{\alpha }) \text{ .}
\end{aligned}
\end{equation}
This answer matches the holomorphic heavy-light semiclassical Virasoro conformal block.

Although we have succeeded in showing how to produce the correct conformal blocks via higher spin gauge transformation, it has to be said that this derivation needs to be understood better.  We already commented above on the singular nature of the bulk-to-boundary propagators that are effectively being employed here, and the question of why this computation is blind to double trace exchanges.   We also needed to set the parameters $\gamma^{(s_1)}=0$ for $m\geq s_p-s_2$ for no very good reason.  It would be good to clarify these issues.

%%%%%%%%%%%%%%%%%%%%%%%%%%%%%%%%%%%%%%%%%%%%%%%%%%%%%%%%%%%%%%%%%%%%%%%%%%%%%%%%
\section{Final comments}
%%%%%%%%%%%%%%%%%%%%%%%%%%%%%%%%%%%%%%%%%%%%%%%%%%%%%%%%%%%%%%%%%%%%%%%%%%%%%%%%

We conclude with a few remarks.

At a purely technical level, one aspect of our scalar field computation that could be improved would be to relax the reality condition on $w$.   This would allow us to cleanly separate the individual chiral blocks from their product.    This is straightforward in principle, but it turns out to be technically challenging to evaluate the resulting integrals in this case. We also mentioned some technical subtleties with our higher spin calculation in the main text.

Moving into more novel territory, our techniques may be combined with gravitational perturbation theory to derive new results away from the strict limits considered so far. For instance, the semiclassical heavy-light Virasoro block is the leading term in a $1/c$ expansion of the exact Virsaoro block expanded around the limit \eqr{hllim}. These $1/c$ corrections can be worked out explicitly in a power series expansion in $1-z$ using Zamolodchikov's recursion relation, or the more efficient recursion relation of \cite{Fitzpatrick:2015zha} adapted to the heavy-light limit specifically. See \cite{Hijano:2015rla} for some explicit results, and \cite{Perlmutter:2015iya} for closed-form, albeit complicated, expressions for coefficients at any order in $1/c$. These results should correspond to quantum fluctuations of the background geometry. It would be interesting to try to reproduce these from a bulk analysis.
 
Similarly, it would also be interesting to see how the simple relation between the global and Virasoro blocks is modified at subleading orders in $1/c$, in the global limit of large $c$ with dimensions fixed. This may be computed in the bulk by incorporating graviton loop corrections to the AdS$_3$ geodesic Witten diagram.

It would be natural to generalize the heavy-light limit to CFTs with $W$-symmetry. Semiclassical $W_N$ conformal blocks for vacuum exchange have been computed in \cite{deBoer:2014sna} with all charges scaling with $c$ in some manner; it would be useful to loosen that requirement.

An important open question in the world of Virasoro blocks is whether there is a compact form for the semiclassical Virasoro block where all operator dimensions scale linearly with $c$. This is the limit usually considered in the context of Liouville theory. Whatever the answer for this block, the expectation  is that its bulk dual  involves a spacetime with interacting conical defects, not unlike a multi-centered black hole solution.  This connection can be seen via the correspondence between Zamolodchikov's monodromy equations and the Einstein equations expressed in Chern-Simons form; see  e.g. \cite{deBoer:2014sna,Hijano:2015rla}. Understanding this picture in detail, and what it implies for various questions in CFT -- e.g.  two-interval R\'enyi entropies \cite{Faulkner:2013yia, Hartman:2013mia} -- would be very interesting.

\vspace{.3in}

\noindent
{ \bf \Large Acknowledgments}

\vspace{.1in}

P.K. is supported in part by NSF grant PHY-1313986. E.P. wishes to thank the KITP, Strings 2015 and the Simons Center for Geometry and Physics for hospitality during this project.  This research was supported in part by the National Science Foundation under Grant No. NSF PHY11-25915. E.P. is supported by the Department of Energy under Grant No. DE-FG02-91ER40671.

\appendix

\section{Evaluation of the geodesic integrals}
\label{integral}

Equation (\ref{result}) of the main text gives an integral expression which reproduces the Virasoro conformal block with an exchanged scalar of conformal dimensions $(h_p,h_p)$. In this appendix we evaluate the integrals and put the result into a form that can be readily compared with the known formula (\ref{FKWw}) for the conformal blocks.

We begin with the integral expression
\es{start}{
I&=\int_{-\infty}^{\infty} d\lambda e^{-\frac{2H_{12}}{\alpha}\lambda}(\cosh\lambda)^{-2h_p}\\ &\times\int_{-\infty}^{\infty} d\lambda' e^{-2h_{12}\lambda'}(\cosh\lambda')^{-2h_p}
{}_2 F_1 \lr{h_p,h_p+\frac{1}{2},2h_p;\frac{(\sin\frac{\alpha w}2)^2}{(\cosh\lambda\cosh\lambda')^2}} \text{ .}}
In terms of which equation (\ref{result}) reads
\begin{align}
\cW_{2h_p,0}(w) = \lr{\sin\tfrac{\alpha w}2}^{2h_p-2h_{L_1}-2h_{L_2}} \times I .
\end{align}
Notice that $I$ receives divergent contributions from large $\lambda$ or $\lambda'$ unless
\begin{align}\label{difsump}
\left| \tfrac{H_{12}}{\alpha} \right| < h_p
 \quad\mathrm{and} \quad
|h_{12}| < h_p.
\end{align}
In what follows we assume  that these conditions are met. A similar assumption was necessary in \cite{Hijano:2015rla}.

We expand the hypergeometric function in powers of $x\equiv \sin^2\frac{\alpha w}2$ to find
\begin{align}\label{Ispans}
I = \sum_{n=0}^{\infty} \lr{\int_{-\infty}^{\infty} d\lambda \,e^{-\frac{2H_{12}}{\alpha}\lambda} (\cosh\lambda)^{-2n-2h_p}}
\lr{\int_{-\infty}^{\infty} d\lambda'\, e^{-2h_{12}\lambda'} (\cosh\lambda')^{-2n-2h_p}} \frac{(h_p)_n(h_p+\tfrac{1}2)_n}{(2h_p)_n n!} x^n  \text{ ,}
\end{align}
where $(h)_n = \frac{\Gamma(h+n)}{\Gamma(h)}$ is the Pochhammer symbol. Condition \eqref{difsump} ensures that both integrals above are finite. They are given by
\begin{align}
\begin{split}
\int_{-\infty}^{\infty} d\lambda \,e^{-\frac{2H_{12}}{\alpha}\lambda} (\cosh\lambda)^{-2n-2h_p} &= 2^{2m-1}B\lr{m-\tfrac{H_{12}}{\alpha},m+\tfrac{H_{12}}{\alpha}}  \text{ ,}\\
\int_{-\infty}^{\infty} d\lambda'\, e^{-2h_{12}\lambda'} (\cosh\lambda')^{-2n-2h_p} &= 2^{2m-1}B\lr{m-h_{12},m+h_{12}}  \text{ ,}
\end{split}
\end{align}
where $B$ is the beta function $B(p,q) = \frac{\Gamma(p)\Gamma(q)}{\Gamma(p+q)}$ . Substituting for the integrals in equation \eqref{Ispans} and then using twice the identity
\begin{align}
\Gamma (2h_p+2n) =
2^{2n}\Gamma (2h_p) (h_p)_n(h_p+\tfrac{1}2)_n  \text{ ,}
\end{align}
which follows from the Legendre duplication formula, we find
\begin{align}
\begin{split}
I = &\frac{2^{4h_p-2}\Gamma (h_p+\tfrac{H_{12}}{\alpha})\Gamma (h_p-\tfrac{H_{12}}{\alpha})\Gamma (h_p+h_{12})\Gamma (h_p-h_{12})}{\Gamma (2h_p)\Gamma (2h_p)}\\
&\times\sum_{n=0}^{\infty}
\frac{(h_p+\tfrac{H_{12}}{\alpha})_n(h_p-\tfrac{H_{12}}{\alpha})_n
(h_p+h_{12})_n(h_p-h_{12})_n }
{ (2h_p)_n (h_p)_n(h_p+\tfrac{1}2)_n n!} x^n  \text{ .}
\end{split}
\end{align}
We recognize the sum on the second line as the power series of a ${}_4 F_3$ hypergeometric function. Let ${\cal N}$ stand for the factor in the top line multiplying this function. Then
\begin{align}\label{Gfin}
\cW_{2h_p,0}(w) = {\cal N}\lr{\sin\tfrac{\alpha w}2}^{2h_p-2h_{L_1}-2h_{L_2}}
{}_4 F_3\left(
\begin{array}{c}
h_p+\tfrac{H_{12}}{\alpha}, h_p-\tfrac{H_{12}}{\alpha},h_p+h_{12},h_p-h_{12} \\
2h_p,\, h_p, \,h_p+\tfrac{1}2
\end{array}
\Big| \sin^2\tfrac{\alpha w}2 \right) \text{ .}
\end{align}
To facilitate comparison with the result (\ref{FKWw}), we would like to write this ${}_4 F_3$ hypergeometric function as a product of ${}_2F_1$ functions. To that end we employ the identity \cite{wolfram}
\begin{align}
\begin{split}\label{4f3ident}
{}_4 F_3\Farg{
a,\,b-a,\,a',\,b-a'}{\frac{b}2,\, \frac{b+1}2, \,b}{\frac{z^2}{4(z-1)}}
= {}_2 F_1\Big(a,a',b;z\Big)\,{}_2 F_1\Big(a,a',b;\frac{z}{z-1}\Big)  \text{ ,}
\end{split}
\end{align}
which is valid when $z\notin \{1,\infty\}$.  Using this identity with
\begin{align}
z=1-e^{i\alpha w} ~,&& a=h_p+h_{12} ~, && a'=h_p-\tfrac{H_{12}}{\alpha} ~, && b= 2h_p~,
\end{align}
one finds
\es{Gfinal}{\cW_{2h_p,0}(w) = {\cal N}\lr{\sin\tfrac{\alpha w}2}^{2h_p-2h_{L_1}-2h_{L_2}}\times\,&{_2{F_1}}\left(h_p+h_{12},h_p-{H_{12}\over \alpha} ,2h_p;1-e^{i\alpha w}\right)\\\times\,&{_2{F_1}}\left(h_p+h_{12},h_p-{H_{12}\over \alpha} ,2h_p;1-e^{-i\alpha w} \right)~,}
which matches (\ref{FKWw}). This is the result \eqr{main} quoted in the main text.

\section{Recovering the worldline approach}
\label{worldapp}

In previous work \cite{Hijano:2015rla} we presented a bulk construction for conformal blocks in a special case of the heavy-light limit \eqr{hllim} considered here. In this appendix we show how that construction arises as a saddle point approximation to the present, more general one.

Specifically, we worked to first order in the limit where $h_{L_1},h_{L_2},h_p$ are large, and in addition assumed $h_{H_1}=h_{H_2}$. In that case we showed the Virasoro conformal partial wave, $W$, to be $W \propto e^{-2S_{\free}}$, where $S_{\free}$ is found by minimizing the action
\e{dxaa}{S = h_{L_1} L_{L_1} + h_{L_2} L_{L_2} + h_p L_p}
of a configuration of worldlines in the conical defect background. The worldlines $L_i$ originate at the external light operators' positions, worldline $p$ originates at the conical defect, and all three meet at a cubic vertex in the bulk. Here $L_p$ is the length of worldline $p$ and $L_{L_i}$ is the length of worldline $L_i$ regularized by putting the boundary points at large but finite distance from the origin.

The subscript ``$\free$'' on $S_{\free}$ is meant to emphasize that the vertex joining worldlines $L_1,L_2,p$ is unconstrained: it will go wherever in the bulk it needs to go in order to make $S$ as small as possible, and in particular it need not lie on the geodesic connecting the light operators' positions.

Meanwhile, in the present approach, setting $h_{H_1}=h_{H_2}$ and taking $h_{L_1},h_{L_2},h_p$ large, the geodesic Witten diagram \eqr{geoblock} becomes
\e{dxa}
{\cW \propto \int d\lambda \int d\lambda' e^{-2S(y(\lambda),y(\lambda'))} \text{ ,}}
where $S(y(\lambda),y(\lambda'))$ is the action of the worldline configuration in which the vertex joining $L_1,L_2,p$ is located at $y(\lambda)$ and the one joining $p$ to the defect is located at $y(\lambda')$. With the light operator dimensions large, $S$ is large, and the leading behavior of the integral in \eqr{dxa} is dominated by the immediate neighborhood of the point $(\lambda,\lambda')$ that minimizes $S$. Therefore
\e{dxb}{\cW \propto e^{-2S_{\geo}[h_{L_1},\,h_{L_2},\,h_p]}~,}
with $S_{\geo}$ found by minimizing the worldline action \eqr{dxaa} with respect to the positions of the two cubic vertices, but now with both vertices constrained to lie on their respective geodesics.

Clearly $S_{\geo}$ and $S_{\free}$ are different (and $S_{\geo}>S_{\free}$). Nevertheless, the two prescriptions $W\propto e^{-S_{\free}}$ and $W\propto e^{-S_{\geo}}$ are in fact the same up to overall normalization, because the difference between $S_{\free}$ and $S_{\geo}$ is a constant, independent of the operator locations, as we will now show.

\subsection{Equivalence of the minimization prescriptions}
Agreement between the prescriptions follows from the following two observations. Here $h$ is some positive number:
\begin{enumerate}[(a)]
\item $S_{\free}[h_{L_1},\,h_{L_2},\,h_p] = S_{\free}[h_{L_1}+h,\,h_{L_2}+h,\,h_p] - S_{\free}[h,\,h,\,0]$, up to a constant.
\item $S_{\geo}[h_{L_1},\,h_{L_2},\,h_p]=S_{\free}[h_{L_1}+h,\,h_{L_2}+h,\,h_p] - S_{\free}[h,\,h,\,0]$ in the limit $h \gg h_{L_1},h_{L_2},h_p$.
\end{enumerate}
It is easy to see that (a) and (b) together imply $S_{\free} = S_{geo}$ up to a constant, as desired.

We work on the cylinder. Property (a) can be read off from the expression obtained in \cite{Hijano:2015rla} for $S_{\free}$ as a function of the separation $w_{12}$ between the light external operators:
\begin{align}
\begin{split}\label{dxd}
S_{\free}[h_{L_1},\,h_{L_2},\,h_p] =&~ (h_{L_1}+h_{L_2})\log\sin{\alpha w_{12}\over 2} + h_p\arctanh\frac{h_p\cos{\alpha w_{12}\over 2}}{\sqrt{h_p^2-(h_{L_2}-h_{L_1})^2\sin^2{\alpha w_{12}\over 2}}}\\
& - |h_{L_2}-h_{L_1}|\log\left(|h_{L_2}-h_{L_1}|\cos\tfrac{\alpha w_{12}}2 + \sqrt{h_p^2-(h_{L_2}-h_{L_1})^2\sin^2\tfrac{\alpha w_{12}}2}\right)\\
&+\text{constant}  \text{ .}
\end{split}
\end{align}
Only the first term and the constant change upon substituting $h_{L_{1,2}}\to h_{L_{1,2}}+h$ and the change in the first term is precisely $S_{\free}[h,\,h,\,0]$.

Proceeding now to prove (b), we start from the fact that when $h$ is much larger than $h_{L_1},h_{L_2},h_p$ the function
\e{dxe}{S = (h_{L_1}+h)L_{L_1} + (h_{L_2} + h)L_{L_2} + h_p L_p}
is minimized when the total length of worldlines $L_1$ and $L_2$ is as small as possible, i.e. when their union is a geodesic. The location of the vertex is then found by minimizing $S$ subject to that constraint. Therefore in the limit $h\gg h_{L_1},h_{L_2},h_p$
\e{dxf}{S_{\free}[h_{L_1}+h,\,h_{L_2}+h,\,h_p] = S_{\geo}[h_{L_1}+h,\,h_{L_2}+h,\,h_p] \text{ .}}
Now, the position of the intersection vertex that gives $S_{\geo}$ depends on the light operator dimensions only through their difference, and a shift of both dimensions by the same amount $h$ merely shifts $S_{\geo}$ by $h(L_{L_1}+L_{L_2})$. Thus equation \eqr{dxf} is equivalent to
\e{dxg}{S_{\free}[h_{L_1}+h,\,h_{L_2}+h,\,h_p] = S_{\geo}[h_{L_1},\,h_{L_2},\,h_p] + h(L_{L_1}+L_{L_2}) \text{ ,}}
and (b) follows from the fact that $S_{\free}[h,\;h,\;0] = h(L_{L_1}+L_{L_2})$.

%%%%%%%%%%%%%%%%%%%%%%%%%%%%%%%%%%%%
\section{Details of some higher spin gravity calculations}\label{ap:Hsap}

In this appendix we fill in some details which are needed for the results in section \ref{higherspin}.  One result we need are the structure constants of the $sl(N,\mathbb{R})\times sl(N,\mathbb{R})$ higher spin algebra. These appear when computing $\delta a^{(0)}$ and $\delta a^{(1)}$ in \ref{eq:da0} and \ref{eq:da1}.

\subsec{$sl(N,\mathbb{R})$ algebra}\label{subsec:slnRalgebra}

The commutator between two generators of spins $s$ and $s^{\prime }$ is
\begin{equation}
\left[ W_{n}^{(s)},W_{m}^{(s^{\prime })}\right] =\sum\limits_{u=2,4,6,...}^{s+s^{\prime }-1}g_{u}^{ss^{\prime }}\left(
n,m;N\right) W_{n+m}^{s+s^{\prime }-u} \text{.}
\end{equation}
The structure constants are denoted by $g_{u}^{ss^{\prime }}\left(n,m;N\right)$ and can be written as a product of a function of $N$ and a function of $m$ and $n$,
\begin{equation}
g_{u}^{ss^{\prime }}\left( n,m;N\right) =\frac{q^{u-2}}{2\Gamma \left(
u\right) }\phi _{u}^{ss^{\prime }}\left( N\right) {\cal N}_{u}^{ss^{\prime }}\left(
n,m\right) \text{ ,}
\end{equation}
where
\begin{equation}\begin{aligned}\label{c3}
\phi_{u}^{ss^{\prime }}\left( N\right)  =&\pFq{4}{3} {\frac{1}{2}-N,\frac{1}{2}+N,\frac{2-u}{2},\frac{1-u}{2}}{\frac{3}{2}-s,\frac{3}{2}-s^{\prime },\frac{1}{2}+s+s^{\prime }-u}{1} \\
{\cal N}_{u}^{ss^{\prime }}\left( n,m\right)  =&\sum\limits_{k=0}^{u-1}\left(
-1\right) ^{k}{{u-1}\choose k}
 \left( 1-s-n\right) _{u-1-k}\left( 1-s+n\right) _{k}
\left( 1-s^{\prime }-m\right) _{k}\left( 1-s^{\prime }+m\right)
_{u-1-k} \text{ ,}
\end{aligned}\end{equation}
$(a)_n=\Gamma(a+n)/\Gamma(a)$ is the rising Pochhammer symbol, and $q$ is a normalization constant that can be scaled away by taking $W_{n}^{(s)}\rightarrow q^{s-2}W_{n}^{(s)}$.

For the purposes of the calculations in the main text, the function $\phi_{u}^{ss^{\prime }}\left( N\right) $ can be ignored, as it does not depend on $m$. This can be seen in the solutions for the functions $\gamma^{(s_1)}_m$ written in equation \ref{eq:gamma}: an $m$-independent number will contribute as a common overall factor to all $\gamma$'s and hence to the conformal block extracted from $J^{(s_1)}(z)$.

A simplification of the structure constants occurs for commutators in which one of the generators is lowest or highest weight, that is, when $n=\pm(s-1)$ or $m=\pm(s'-1)$. This is clear from \eqr{c3}, using the fact that $(0)_n=\delta_{n,0}$. When $n=-(s-1)$, say, only the $k=u-1$ term survives the sum. Using  also the  identity $(-x)_{n}=(-1)^n (x-n+1)_n$, we can write
\begin{equation}\begin{aligned}\label{eq:struccons}
g_{u}^{s s^{\prime}}\left( -s+1,m;N\right) \sim  (-1)^{u-1} \left( 2s-u \right)_{u-1}  \left( m+s^{\prime}-u+1 \right)_{u-1} \text{ ,}
\end{aligned}
\end{equation}
(We use $\sim$ to mean that we drop prefactors that do not affect our arguments, like $\phi^{ss'}_u(N)$.) These structure constants appear in our equation \ref{eq:da1}, for which $s=s_2$, $n=-(s_2-1)$, and $s^{\prime}=s_p$.  Defining the spin of the resulting generator as $s_{1}=s_{2}+s_{p}-u$, we then have
\begin{equation}\label{eq:struccons}
g_{s_p-s_{12}}^{s_2 s_p}\left( -s_2+1,m;N\right) \sim(-1)^{s_p-s_{12}-1} \left( s_1+s_2-s_p \right)_{s_p-s_{12}-1}  \left( m+s_{12}+1 \right)_{s_p-s_{12}-1} \text{ .}
\end{equation}
After the replacement $m=i+j-1-s_{12}$ and ignoring the factors that do not depend on $i$ or $j$, equation (\ref{eq:struccons}) matches formula (\ref{eq:simpleg}) used to compute  $J^{(s_1)}(z)$ in (\ref{eq:J}).

\subsec{Deriving \eqr{eq:Js1}}

We proceed now to derive the expression for $J^{(s_1)}(z)$ in \eqr{eq:Js1} starting from equation (\ref{eq:J}), where we have made use of the simplification of the structure constants in equation (\ref{eq:struccons}). The starting point is
\begin{equation}\begin{aligned}
J^{(s_{1})}\left( z\right)  \sim
&\sum\limits_{i=0}^{s_{p}+s_{12}-1}\sum%
\limits_{j=1}^{s_{p}+s_{12}-i}\frac{\Gamma \left(
i+j+s_{p}-s_{12}-1\right) \Gamma \left( 2s_{1}-i-j\right) \Gamma \left(
2s_{2}+i\right) }{\Gamma \left( i+1\right) \Gamma \left( j\right) } \\
&\times \left( 1-z\right) ^{-2s_{2}-i}\left( -1\right) ^{j}\partial
_{z}^{j-1}y_{i+j-1-s_{12}}^{(s_{p})} \text{ .}
\end{aligned}\end{equation}
The parameters $y_{m}^{(s_{p})}$ can be replaced by derivatives of $y_{s_{p}-1}^{(s_{p})}$ in virtue of equation \ref{eq:RecYm}. At this point the sum to evaluate  reads
\begin{equation}\begin{aligned}
J^{(s_{1})}\left( z\right)  \sim &\sum\limits_{i=0}^{s_{p}+s_{12}-1}%
\frac{\Gamma \left( 2s_{2}+i\right) }{\Gamma \left( i+1\right) }%
 \left( 1-z\right) ^{-2s_{2}-i}\left( -1\right) ^{i}\partial
_{z}^{-i+s_{p}+s_{12}-1}y_{s_{p}-1}^{(s_{p})} \\
&\times
\sum\limits_{j=1}^{s_{p}-s_{2}+s_{1}-i}\frac{\Gamma \left(
i+j+s_{p}-s_{12}-1\right) \Gamma \left( i+j-2s_1+1\right) }{\Gamma
\left( i+j-s_{12}-s_p\right) \Gamma \left( j\right) }
 \text{ .}
\end{aligned}\end{equation}
The sum over $j$ does not involve any dependence on $z$ and can now be performed analytically. Because the denominator diverges for $j>s_p+s_{12}-1$, the sum over $j$ is a $_2 F_1$ hypergeometric function of argument $1$. Applying Gauss's hypergeometric theorem \cite{wolframGHT} leads to the following equation for $J^{(s_1)}(z)$
\begin{equation}\begin{aligned}
J^{(s_{1})}\left( z\right)  \sim \sum\limits_{i=0}^{s_{p}+s_{12}-1}%
\frac{\Gamma \left( 1+i-s_{12}-s_{p}\right) \Gamma \left(
i-s_{12}+s_{p}\right) }{\Gamma \left( i+1\right) } {{ \partial
_{z}^{-i+s_{p}+s_{12}-1}y_{s_{p}-1}^{(s_{p})} }\over{(z-1)^{2s_2+i}}}\text{ .}
\end{aligned}\end{equation}
The next step is to replace the derivatives of $y_{s_{p}-1}^{(s_{p})}$
by hypergeometric functions using \ref{eq:Ysm1}. After some algebra we can write
\begin{equation}\begin{aligned}
J^{(s_{1})}\left( z\right)  \sim &\left( 1-z\right)
^{s_{p}-s_{2}-s_{1}}\sum\limits_{i=0}^{s_{p}+s_{12}-1}{{\Gamma\left( 1+i-s_{12}-s_p  \right)}\over{\Gamma\left( 1-s_{12}-s_p \right)}}\frac{ \Gamma\left( i+s_p-s_{12} \right) }{\Gamma \left( i+1\right) \Gamma \left( 1+
i+s_{p}-s_{12}\right) } \\
&\times \text{ }{ _2 F_1}\left(1,s_{p}-S_{12}; s_{p}-s_{12}+i+1;1-z\right)  \text{ .}
\end{aligned}\end{equation}
Invoking the definition of the hypergeometric function as a sum over an integer $r$, one can perform the sum over $i$. The relevant sum is
\begin{equation}\begin{aligned}\label{eq:sumi}
\sum\limits_{i=0}^{s_{p}+s_{12}-1}
{{\Gamma\left( 1+i-s_{12}-s_p  \right)}\over{\Gamma\left( 1-s_{12}-s_p \right)}}\frac{ \Gamma\left( i+s_p-s_{12} \right) }{\Gamma \left( i+1\right) \Gamma \left( 1+
i+s_{p}-s_{12}+r\right) }
\sim \frac{\left( s_{p}+s_{12}\right) _{r}}{\left( 2s_{p}\right)
_{r}\Gamma \left( r+1\right) } \text{ ,}
\end{aligned}\end{equation}
where $r$-independent overall factors have been ignored. In \ref{eq:sumi} we first note that we can let $i$ run from $0$ to $\infty$ because for $i>s_p+s_{12}-1$ each contribution to the sum vanishes. We can then identify the sum as a ${}_2 F_1$ hypergeometric function of argument $1$ and use Gauss's hypergeometric theorem to obtain the expression in \ref{eq:sumi}. The final answer can be resummed into a $z$ dependent hypergeometric function
\begin{equation}\begin{aligned}
J^{(s_{1})}\left( z\right)  \sim &\left( 1-z\right)
^{s_{p}-s_{1}-s_{2}}\sum\limits_{r=0}^{\infty }\text{ }\frac{\left(
s_{p}-S_{12}\right) _{r}\left( s_{p}+s_{12}\right) _{r}}{\left(
2s_{p}\right) _{r}\Gamma \left( r+1\right) }\left( 1-z\right) ^{r} \\
=&\left( 1-z\right)^{s_{p}-s_{1}-s_{2}}{ _2 F_1}\left(s_{p}-S_{12},s_{p}+s_{12};2s_{p};1-z\right)  \text{ .}
\end{aligned}\end{equation}
This result is also written in equation \ref{eq:Js1} of section \ref{subsec:calchs}.

\bibliographystyle{ssg}
\bibliography{biblio}

\end{document}